\def\isreviewversion{0} 

\PassOptionsToPackage{breaklinks}{hyperref}

\if 1\isreviewversion
  \documentclass[manuscript, review]{acmart-authorversion}
\else 
  \documentclass[sigconf]{acmart}
\fi

\copyrightyear{2021}
\acmYear{2021}
\setcopyright{acmlicensed}
\acmConference[CHI '21]{CHI Conference on Human Factors in Computing Systems}{May 8--13, 2021}{Yokohama, Japan}
\acmBooktitle{CHI Conference on Human Factors in Computing Systems (CHI '21), May 8--13, 2021, Yokohama, Japan}
\acmPrice{15.00}
\acmDOI{10.1145/3411764.3445421}
\acmISBN{978-1-4503-8096-6/21/05}

\acmSubmissionID{2068}

\usepackage{booktabs} 
\usepackage{stfloats}
\usepackage{soul} 
\usepackage{multirow}
\usepackage{array}
\usepackage{color}
\usepackage{colortbl}
\usepackage{longfbox} 
\usepackage{subcaption}

\newfboxstyle{patternparam}{padding=1.5pt, margin-bottom=1pt, margin-top=1pt, border-radius=3pt, background-color=tableheader, border-style=none, height=6pt}

\newcommand{\thicktablehline}{\specialrule{0.75pt}{0pt}{0pt}}

\renewenvironment{quote}
  {\list{}{\rightmargin=0.25cm \leftmargin=0.25cm}%
   \item\relax}
  {\endlist}

\newcolumntype{P}[1]{>{\centering\arraybackslash}p{#1}}

\newcommand{\weblink}{\urlstyle{tt}\url{https://data-at-hand.github.io}}

\newcommand{\weblinksupple}{\urlstyle{tt}\url{https://data-at-hand.github.io/chi2021}}

\definecolor{youngho}{RGB}{27,158,119}
\definecolor{bongshin}{RGB}{217,95,2}
\definecolor{eunkyoung}{RGB}{102,166,30}
\definecolor{arjun}{RGB}{20, 201, 192}

\definecolor{revised}{RGB}{0,0,255}

\if 1\isreviewversion
    \newcommand{\revised}[1]{\textcolor{revised}{#1}}
\else
    \newcommand{\revised}[1]{#1}
\fi

\if 1\isreviewversion
    \newenvironment{revisedblock}{\color{revised}}{}
\else
    \newenvironment{revisedblock}{}{}
\fi

\definecolor{tableheader}{HTML}{EFEFEF}
\definecolor{tablegrayline}{HTML}{d0d0d0}

\newcommand{\symbolmic}{\raisebox{-1.5pt}{\includegraphics[width=10pt]{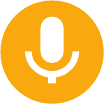}} }
\newcommand{\symbolmictiny}{\raisebox{-2pt}{\includegraphics[width=8.5pt]{figures/glyph_mic.pdf}} }

\newcommand{\symbolmulttiny}{\raisebox{-2.5pt}{\includegraphics[width=9pt]{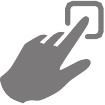}} }

\newcommand{\ipstart}[1]{\vspace{1mm} \noindent{\textbf{\textit{#1.}}}}

\newcommand{\bnpstart}[1]{\vspace{1mm} \noindent{\textbf{#1}}}

\newcommand{\autorefsuffix}[2]{\hyperref[#1]{\autoref{#1}#2}}

\newcommand{\circledigit}[1]{\textbf{\normalsize{\textsf{\textcircled{\footnotesize{#1}}}}}}

\newcommand{\chartplotname}{aggregation plot}

\newcommand\participantcolumnwidth{0.024\textwidth}
\newcommand{\patterncode}[1]{\textbf{\scriptsize{#1}}}

\newcommand{\patternexample}[1]{{\color{lightgray}\rule{3cm}{0.4pt}}\\ #1 \vspace{0.2mm}}

\definecolor{patterntime}{HTML}{ebc0ce}
\definecolor{patterntimetext}{HTML}{d1567f}

\definecolor{patterndatasource}{HTML}{e6f2e6}
\definecolor{patterndatasourcetext}{HTML}{328732}

\definecolor{patterncycle}{HTML}{d5eef2}
\definecolor{patterncycletext}{HTML}{2ebbd1}

\definecolor{patterncondition}{HTML}{f5e9d0}
\definecolor{patternconditiontext}{HTML}{bd7004}

\definecolor{indirect}{HTML}{757575}

\newcommand{\labelphantom}[1]{%
  \parbox{0pt}{\phantomsubcaption\label{#1}}%
}

\begin{document}

\title{Data@Hand: Fostering Visual Exploration of Personal Data on~Smartphones Leveraging Speech and Touch Interaction}


    \author{Young-Ho Kim}
    \affiliation{%
          \institution{University of Maryland}
          \city{College Park}
          \state{MD}
          \country{USA}
    }
    \email{yghokim@umd.edu}
    
    \author{Bongshin Lee}
    \affiliation{%
          \institution{Microsoft Research}
          \city{Redmond}
          \state{WA}
          \country{USA}
    }
    \email{bongshin@microsoft.com}
    
    \author{Arjun Srinivasan}
    \authornote{Arjun Srinivasan conducted this work while with Georgia Institute of Technology.}
    \affiliation{%
        \institution{Tableau Research}
        \city{Seattle}
        \state{WA}
        \country{USA}
    }
    \email{arjunsrinivasan@tableau.com}
    
    \author{Eun Kyoung Choe}
    \affiliation{%
        \institution{University of Maryland}
        \city{College Park}
        \state{MD}
        \country{USA}
    }
    \email{choe@umd.edu}


\renewcommand{\shortauthors}{Young-Ho Kim, Bongshin Lee, Arjun Srinivasan, and Eun Kyoung Choe}

\begin{abstract}
Most mobile health apps employ data visualization to help people view their health and activity data, but these apps provide limited support for visual data exploration. Furthermore, despite its huge potential benefits, mobile visualization research in the personal data context is sparse. This work aims to empower people to easily navigate and compare their personal health data on smartphones by enabling flexible time manipulation with speech. We designed and developed Data@Hand, a mobile app that leverages the synergy of two complementary modalities: speech and touch. Through an exploratory study with 13 long-term Fitbit users, we examined how multimodal interaction helps participants explore their own health data. Participants successfully adopted multimodal interaction (i.e., speech and touch) for convenient and fluid data exploration. Based on the quantitative and qualitative findings, we discuss design implications and opportunities with multimodal interaction for better supporting visual data exploration on mobile devices.
\end{abstract}



\begin{teaserfigure}
    \includegraphics[width=\textwidth]{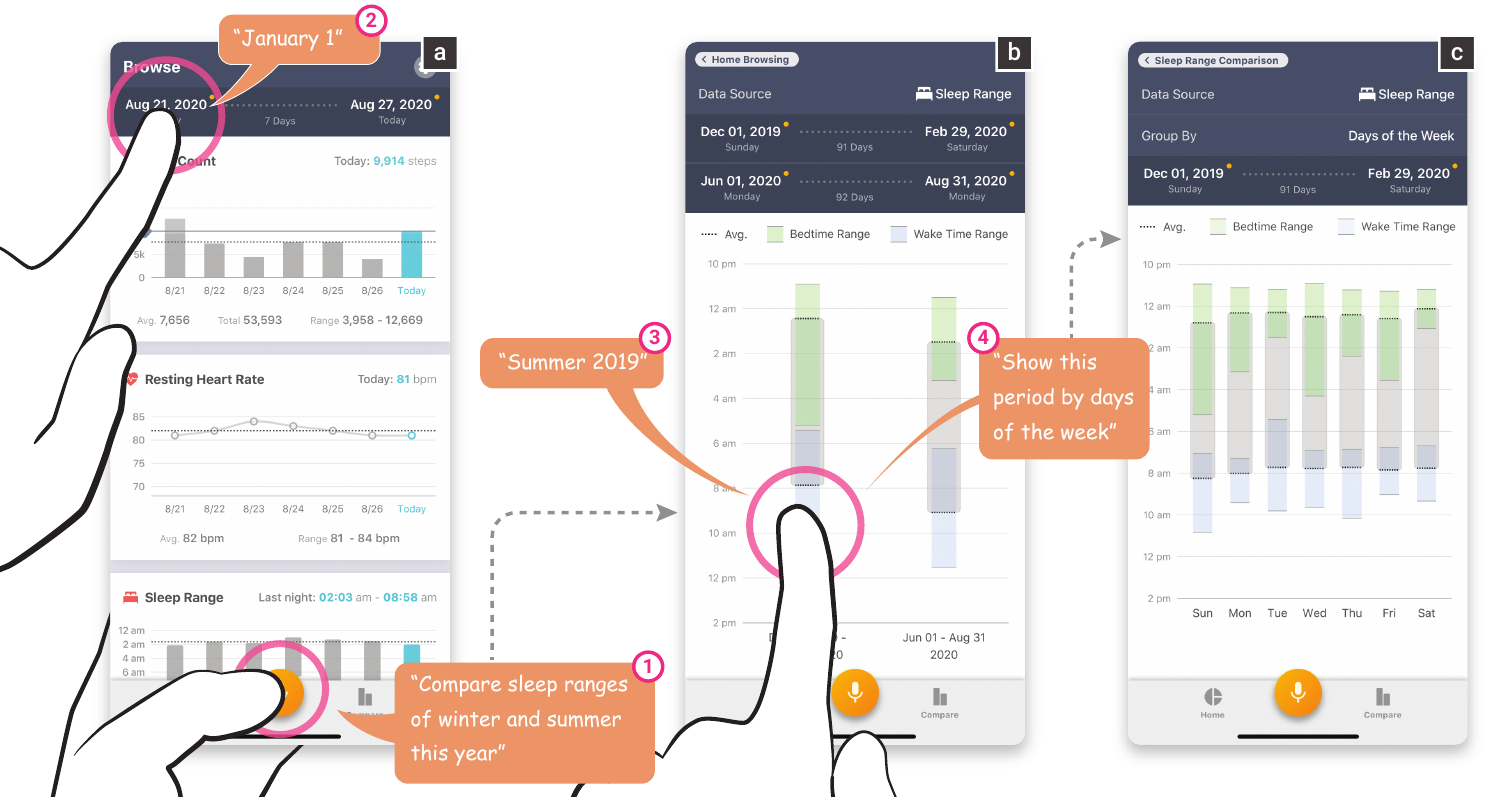}
    \labelphantom{fig:teaser:home}
    \labelphantom{fig:teaser:tworange}
    \labelphantom{fig:teaser:cyclical}
    \vspace{-3mm}
    \caption{Data@Hand supports multimodal interactions to enable people to easily navigate and compare their personal health data on smartphones. People can execute a context-agnostic command such as setting up a comparison by specifying two new periods using a global speech button~\circledigit{1}. They can feed a context to their utterance by touch, such as the start date~\circledigit{2}, the target for comparison~\circledigit{3}, or the time range for refining the view~\circledigit{4}.
    (Please refer to our supplementary video, available at \weblinksupple, which demonstrates the interactions.)
    }
    \Description{Screenshots of Data@Hand, explaining core interactions with hand images.}
    \label{fig:teaser}
    \vspace{3mm}
\end{teaserfigure} 

\maketitle

\section{Introduction}
\label{sec:introduction}

Smartphones, equipped with high-resolution displays and powerful processors, are increasingly becoming a dominant way to access information~\cite{DeviceUsage}.
A vast number of mobile health (or mHealth) apps, including wearable devices' companion apps (e.g., Fitbit App~\cite{Fitbit}, Apple Health~\cite{AppleHealth}, Samsung Health~\cite{SHealth}, Garmin~\cite{garmin}, and Mi Fit~\cite{MiFit}), enable people to access their health and activity data collected over time.
While these mHealth apps commonly employ data visualizations to help people view and understand personal data~\cite{alrehiely2018taxonomy}, they provide limited support for navigating and exploring the data.
Furthermore, research on mobile data visualization is sparse~\cite{Lee2018MobileVisualization}. Much of the mobile visualization research has been carried out with tablets~\cite{Blumenstein2016EvaluatingMobileInfoVis}, and only a handful of projects have recently begun to study data visualizations on smartphones (e.g.,~\cite{brehmer2018visualizing,schwab2019evaluating,brehmer2019comparative}) and smartwatches (e.g.,~\cite{blascheck2018glanceable,klamka2020watch+,Chen2017TimeSeriesSmartwatch}).

In this work, we investigate how to facilitate flexible data exploration on smartphones in the context of self-tracking data, while addressing several challenges smartphones pose.
Due to their limited screen space, smartphones cannot afford a control panel of widgets (alongside the visualizations), which are effective means to support dynamic queries~\cite{ahlberg2003visual}. It is distracting to navigate to a separate page to adjust the widgets and come back to the page with visualizations to see the effect.
In addition, the lack of mouse input makes it difficult to perform two essential actions---(1) a precise selection and (2) details-on-demand using a mouse hover interaction---which are well supported in a desktop environment.
Furthermore, while \textit{time} is a primary dimension of self-tracking data, it is laborious to perform time-based interactions on smartphones, such as entering specific date, time, and ranges. As a result, most mHealth apps tend to limit time manipulations. For example, the Fitbit App restricts people to view data by predefined time segments, such as one week, one month, three months, and one year.
Inspired by previous research advocating the benefits of multimodal interaction~\cite{Choe2018SpeechOnMobileDevice, Lee2018MultimodalDataVisWorkshop, Lee2020BroadAudienceVis, Turk2014MultimodalInteraction, Coutaz1995UsabilityMultimodal}, we incorporate an additional input modality, \textit{speech}, to overcome these challenges. 
Speech-based interaction takes little screen space. Speech is flexible to cover different ways that people specify date (e.g., ``Last Thanksgiving'' or ``Lunar New Year's Day'') and date ranges (e.g., ``2017'' indicating the range from January 1, 2017 to December 31, 2017), which people are already familiar with. 

Combining two complementary modalities, speech and touch, we designed and developed Data@Hand (\autoref{fig:teaser}), a mobile app that facilitates visual data exploration. 
As a first step, informed by prior work on personal insights~\cite{Choe2015VisualizationInsights,Choe2017VisualizedSelf}, we support navigation and temporal comparisons of personal health data, as well as data-driven queries. 
To understand how speech and touch interaction can help lay individuals explore their data, we conducted an exploratory study with 13 long-term Fitbit users using Data@Hand. 

We observed that participants successfully adopted multimodal interaction, using both speech and touch interactions while finding personal insights.
Participants reported that they made deliberate choices between the two input modalities for a more convenient and fluid data exploration.
Flexible time expressions enabled by speech-based natural language interaction helped them freely navigate data in a specific time frame (e.g., \symbolmic ``\textit{Go to March 2020}''), quickly set up comparisons (e.g., \symbolmic ``\textit{Compare sleep ranges of winter and summer this year,}'' illustrated in \circledigit{1} in \autoref{fig:teaser}), and easily execute data-driven queries (e.g., \symbolmic ``\textit{Days I walked more than 10,000 steps last month}''). 
Speech commands combined with touch input (e.g., \circledigit{2}, \circledigit{3}, and \circledigit{4}~in \autoref{fig:teaser}) enabled easy modifications of the time components. For example, to change the start/end date, one can simply utter a specific date while holding on the start/end date label.
Also, graphical widgets (e.g., calendar widget, data source drop-down list) served as a fallback to correct erroneous results of speech or to explore a set of categorical values.
Being satisfied with their overall experiences, all but one participant expressed that they are willing to keep using Data@Hand after the study.
The key contributions of this work are:

\bnpstart{(1)} The design and implementation of Data@Hand, the first mobile app that leverages the synergy of speech and touch input modalities for personal data exploration. Data@Hand helps people interact with their own personal data on smartphones by accessing the Fitbit data using the Fitbit REST API. It runs on both iOS and Android, using the Apple speech framework~\cite{AppleSiriHealth} and Microsoft Cognitive Speech API~\cite{MicrosoftCognitiveSpeech} as speech recognizers. \revised{The Data@Hand source code is available at \weblink.}

\bnpstart{(2)} An empirical study conducted with 13 long-term Fitbit users using Data@Hand. From the quantitative and qualitative analysis, we provide an understanding of how people explore their own data using speech and touch interaction on smartphones, uncovering situations and rationale for people's choice of interaction.

\bnpstart{(3)} Design implications and opportunities for multimodal interaction for mobile data visualization. Reflecting on our observations and participants' feedback, we draw design implications and opportunities for developing a multimodal interaction to better support personal data exploration on mobile devices. 
\section{Related Work}
In this section, we cover related work in the areas of (1) visual exploration of personal data and (2) natural language and multimodal interaction for visual data exploration.

\subsection{Visual Exploration of Personal Data}
\label{sec:VDA}




As collecting and reflecting on personal data has become commonplace, research on personal visualization has gained increasing attention~\cite{Li2011SelfReflection,huang2014personal,Lee2020BroadAudienceVis,thudt2017expanding}. 
Personal visualizations equipped with interactivity enable \emph{visual data exploration}, making it easy for people to understand and reflect on their data. 
As such, many Personal Informatics systems (e.g.,~\cite{Aseniero2020ActivityRiver,Choe2017VisualizedSelf,epstein2014taming,Thudt2016VisualMementos,feustel2018people,Huang2016CalendarVis}) support visual data exploration to empower people to gain personal insights.

Because \emph{time} is a primary dimension of personal data (or self-tracking data), systems that support visual data exploration strive to enable easy manipulation of the time component. For example, Visualized Self~\cite{Choe2017VisualizedSelf}---a web application that enables people to integrate and explore personal data from multiple self-tracking services---employs the timeline mini-map to enable rapid adjustment of the data scope. Activity River~\cite{Aseniero2020ActivityRiver} takes a similar timeline-based approach, but the scope was fixed as a single day. Visual Mementos~\cite{Thudt2016VisualMementos} supports visual exploration of personal location history. It incorporates a multidimensional selection widget for the precise scoping of event episodes in a series of location logs. 
Huang and colleagues~\cite{Huang2016CalendarVis} designed an on-calendar visualization tool that integrated people's physical activity data. They chose to leverage a calendar, an inherent time-based visualization with rich personal context, to make the data readily accessible.
We note that these systems were designed for a desktop environment and did not investigate how their interfaces could be applicable in a mobile environment with smaller screen space.

Commercial mHealth apps, including wearable devices' companion apps, also provide the visual exploration capability. However, most of these apps are limited in terms of time navigation, making it hard to jump to an arbitrary time frame or to compare data from two different time frames.
These commercial apps usually show daily information using a dashboard on their main page, aiming to promote self-awareness of the current performance. 
They are also constrained by the smartphone form factor, such as small screen and imprecise touch input. 
Furthermore, existing widgets (e.g., calendar) for date entry are not flexible enough to handle the various ways to specify time.
As a result, limited navigation support becomes a barrier to performing flexible data exploration, and in turn obtaining personal data insights on smartphones.

The practitioner community developed ample applications of data visualization in mobile apps \& websites (refer to~\cite{SadowskiMobileVis, RosMobileVis} for curated practices). Data visualization is also commonly used for mobile form factors, such as smartphones and tablets, in research prototypes (e.g.,~\cite{Choe2015SleepTight,desai2018pictures,Kay2012Lullaby,kay2016ish,pina2020dreamcatcher,schneider2019communicating}) developed by UbiComp and Human-Computer Interaction researchers.  
However, research specifically focusing on mobile data visualization is sparse and much of the mobile visualization research has been carried out with tablets~\cite{Blumenstein2016EvaluatingMobileInfoVis}.
As such, the research community has recently put efforts to shape a research agenda for mobile data visualization while calling for more research endeavors~\cite{Lee2018MobileVisualization,choe2019mobile,Lee2020BroadAudienceVis}. 
Although only a handful, mobile visualization research has begun to pay attention to the smaller form factors (i.e., smartphones and smartwatches). They examined effectiveness of visual representations (e.g., ranges on timeline~\cite{brehmer2018visualizing}, animated transition vs. small multiples of scatterplots~\cite{brehmer2019comparative}, data comparison on smartwatches~\cite{blascheck2018glanceable}) and interaction techniques (e.g., multivariate network exploration~\cite{Eichmann2020Orchard}, pan and zoom timelines and sliders~\cite{schwab2019evaluating}). In addition, in their workshop paper, Choe and colleagues~\cite{Choe2018SpeechOnMobileDevice} envisioned a scenario where speech interaction could facilitate personal data exploration on mobile devices.
Inspired by this line of research and vision, we contribute to mobile data visualization with Data@Hand, the first mobile app that leverages the synergy of speech and touch input modalities to augment personal data exploration.

\subsection{Natural Language and Multimodal Interaction for Visual Data Exploration}

Advancements in natural language understanding and speech recognition technology have promoted the design of natural language interfaces (NLIs) for data visualization~\cite{Srinivasan2018Orko, Srinivasan2020InChorus, Srinivasan2020DataBreeze, Setlur2016Eviza, Dhamdhere2017Analyza, Kassel2018Valletto, Gao2015DataTone, Kassel2019AnswerSpaceVisualAnalysis, Yu2020FlowSense, Cox2001InformationDistillery, Hearst2019ChartConversational, Kumar2017Articulate2, Aurisano2016Articulate2,sun2010articulate,hoque2017applying,setlur2019inferencing}.
These systems commonly exploit two advantages of natural language:~(1)~high flexibility in synthesizing multiple commands and parameters~\cite{Cox2001InformationDistillery} and (2)~low barriers in expressing intents and questions regarding the data~\cite{Srinivasan2017NaturalLangaugeInterface, Aurisano2015ShowMeData}.
A majority of these systems are primarily designed for desktop settings and investigate typed natural language input (e.g.,~\cite{Cox2001InformationDistillery,sun2010articulate,Setlur2016Eviza,Gao2015DataTone,hoque2017applying,Dhamdhere2017Analyza}).
On the other hand, another subset of prior systems focus on speech input and explore multimodal visualization interfaces that incorporate speech with other modalities such as pen and/or touch in post-WIMP~\cite{VanDam2001PostWIMP} settings including tablets~\cite{Kassel2018Valletto,Srinivasan2020InChorus} and large displays~\cite{Srinivasan2018Orko,Srinivasan2020DataBreeze,Kumar2017Articulate2}.

Given the context of the smartphone form factor, two existing systems that are most relevant to our work are the tablet-based systems, Valletto~\cite{Kassel2018Valletto} and InChorus~\cite{Srinivasan2020InChorus}. 
Valletto allows people to specify charts through speech and then perform simple touch gestures such as a rotate to flip axes and swipe to change the visualization type.
Exploiting the complementary nature of pen/touch and speech~\cite{cohen1989synergistic}, InChorus illustrates a vocabulary of multimodal actions involving a wide range of visualizations for tabular data (e.g., selecting marks with a pen and saying ``\textit{Remove others}" to filter unselected points, pointing on an axis with a finger and speaking attribute names to specify data mappings).

Our work extends this line of research on multimodal visualization interfaces on mobile devices in two notable ways.
First, unlike Valletto and InChorus that were designed for tablets, Data@Hand is the first system specifically designed for smartphones.
Correspondingly, we discuss the unique constraints we faced in the form of more limited screen space and lower precision with touch input~\cite{Watson2015MobileVisTutorial}, and how these constraints impacted our interface and interaction design (see \hyperref[sec:dr3]{DR3} in Section ~\ref{sec:dr}).
Second, compared to prior NLIs that primarily target avid users of visualization tools (e.g., data analysts, developers or managers in visualization-oriented products), our work targets a broader population of lay individuals interested in exploring their personal health data.
In doing so, we highlight the role of lay individuals that affected the design of Data@Hand, and discuss in detail the interaction patterns individuals employed for performing tasks such as time manipulation, temporal comparisons, and data-driven queries with Data@Hand.

\section{Data@Hand}
To enable flexible data exploration on smartphones, we designed and developed Data@Hand, a mobile app that leverages the benefits of two complementary input modalities, speech and touch.
In this section, we describe our design rationales and the Data@Hand app along with the implementation details.

\begin{figure*}[b]
    \centering
    \includegraphics[width=\textwidth]{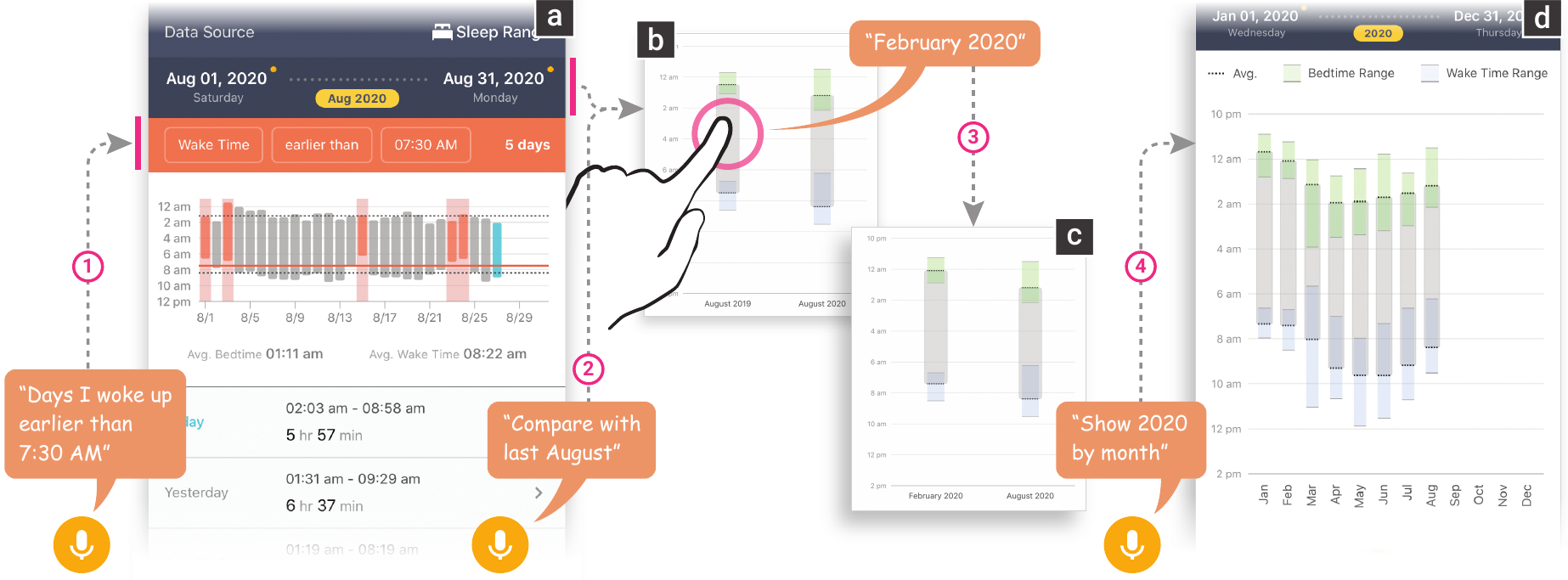}
    \labelphantom{fig:scenario:detail}
    \labelphantom{fig:scenario:tworange:aug}
    \labelphantom{fig:scenario:tworange:feb}
    \labelphantom{fig:scenario:cyclical}
    \vspace{-3mm}
    \caption{Excerpt of the exploration flow in our usage scenario. \circledigit{1}~Zoe executes a data-driven query via natural language. \circledigit{2}~The system infers omitted parameters (e.g., a pre-selected date range as a comparison target) using the current screen information. \circledigit{3}~Zoe changes August 2019 to February 2020 by touch+speech interaction on the \chartplotname{}. \circledigit{4}~Zoe establishes the cyclical comparison to see this year's monthly trend.
    }
    \Description{The exploration flow of Data@Hand, with a series of screenshots.}
    \label{fig:scenario}
\end{figure*}

\subsection{Design Rationales}
\label{sec:dr}

\phantomsection{}
\label{sec:dr1}
\ipstart{DR1: Use Simple and Familiar Visualizations to Support Lay Individuals}
Our target audience is people who collect their health and activity data using commercial wearable devices. They usually do not have expertise in data visualization and analytics. Therefore, we incorporated familiar visual representations and conventional charts that people commonly encounter on existing mHealth apps. For example, we used bar charts, line charts, and range charts to visualize daily measurement values (e.g., step counts and resting heart rates in \autoref{fig:teaser:home} and sleep ranges in \autoref{fig:scenario:detail}). As for visualizing the aggregated data over a period, we designed custom representations called \textit{\chartplotname{}s}: a simplified version of box-and-whisker plots. We encoded only the average and the minimum and the maximum values (for the range) because these metrics are more relevant to the personal tracking context than median and percentiles. For example, \autoref{fig:scenario:tworange:feb} shows the aggregated sleep ranges for February and August of 2020.

\phantomsection{}
\label{sec:dr2}
\vspace{1mm}
\ipstart{DR2: Enable Flexible Time Manipulation to Help Identify Personal Insights}
Unlike the general and broader visual data exploration, personal visualizations require different design considerations because of the nature of the personal data and the diverse personal data collection goals~\cite{huang2014personal,thudt2017expanding}. In visual exploration with personal data (or self-tracking data), people look for specific personal insights, such as whether they achieved a certain personal goal (e.g., 10,000 steps per day), how their behaviors (e.g., step counts, sleep patterns) and emotional/physiological states (e.g., mood, heart rates) change over time, and what factors might have affected the changes (e.g., before \& after the COVID-19 lockdown) ~\cite{Choe2017VisualizedSelf}. 
Furthermore, as shown in prior works~\cite{Choe2015VisualizationInsights, Choe2017VisualizedSelf, Li2011SelfReflection, Sukumar2020TesseraeExplorer}, \textit{comparison by time segmentation} is one of the most common visual exploration tasks that people actively perform to gain personal insights. 
However, flexible time manipulation---a key facilitator in drawing personal insights---is rarely supported in mHealth apps due to the limitations we described earlier (see Section~\ref{sec:VDA}). 
We strove to enable flexible time manipulation, focusing on navigation (\autoref{fig:teaser:home}) and temporal comparisons (Figure \ref{fig:teaser:tworange} \& \ref{fig:teaser:cyclical} and Figure \ref{fig:scenario:tworange:feb} \& \ref{fig:scenario:cyclical}).

\phantomsection{}
\label{sec:dr3}
\vspace{1mm}
\ipstart{DR3: Leverage the Synergy of Speech and Touch Interactions on Smartphones}
Smartphones are increasingly becoming a dominant way to access information~\cite{DeviceUsage}, and much of the personal data is collected from smartphones and wearable devices. As such, we wanted to facilitate easy access to people's own data on smartphones.
To overcome the challenges smartphones pose, we leverage both speech and touch input modalities that are complementary in nature: speech input affords a high \textit{freedom of expression} without requiring much screen space, whereas touch input supports \textit{direct} interaction~\cite{Srinivasan2020InChorus, Srinivasan2018Orko, Srinivasan2020DataBreeze, Saktheeswaran2020OrkoTouchOrSpeech}. 
When combining the strengths of these two modalities, we aimed to provide a complementary set of operations~\cite{Coutaz1995UsabilityMultimodal} rather than providing an equivalent set of operations for each modality. We detail how we synergistically incorporated the two modalities in supporting a diverse set of operations on smartphones in Section~\ref{sec:mm-interaction}.

\subsection{User Interface and Interaction Design} 
Data@Hand currently supports five health metrics that are retrieved from the Fitbit data sources: (1) step count, (2) resting heart rate, (3) sleep duration, (4) hours slept, and (5) weight. For sleep, we included only one sleep log per day that is marked as the main sleep. We chose the metrics based on their prevalence in commercial health apps according to a recent survey~\cite{Kim2019HealthDataAccessibility}.

\subsubsection{User Interface and Interaction Components}
Data@Hand supports navigation, temporal comparison, and data-driven queries using four main pages---\textit{Home}~(\autoref{fig:teaser:home}), \textit{Data Source Detail}~(\autoref{fig:scenario:detail}), \textit{Two-range Comparison}~(\autoref{fig:teaser:tworange}), and \textit{Cyclical Comparison}~(\autoref{fig:teaser:cyclical}). As a default view, the Home page visualizes the past 7-day data for the five data sources. The Data Source Detail page shows detailed information for a single data source. Both comparison pages juxtapose aggregated measurement values, but in two different ways: the Two-range Comparison page plots two selected periods side by side, and the Cyclical Comparison page displays values within a specific period grouped by a predefined time cycle, such as days of the week (i.e., Sunday through Saturday) and months of the year (i.e., January through December).

These pages contain common interaction components. 
An app header and bottom toolbar are located on all pages (see \autoref{fig:teaser}).
The \textit{range widget} on the app header is for manipulating date ranges, and  
the global microphone button~\symbolmic on the toolbar is used to execute a speech command in a global scope---a command that is agnostic to the current context or page. We describe its functionalities in more detail in Section~\ref{sec:mm-interaction}. The Home button is a shortcut that brings back to the Home page, maintaining the current date range. The Compare button opens the configuration panel where people can configure data source, comparison type, and date ranges to execute comparison queries.

People can execute \textit{data-driven queries} by specifying a condition in natural language. 
Data@Hand responds by highlighting the data items in red (see \autoref{fig:scenario:detail}).
Also, the \textit{query bar} (\circledigit{1} in \autoref{fig:scenario:detail}), shown below the app header, contains \textit{parameter widgets} for manipulating recognized parameters in a query (e.g., [Wake Time], [earlier than], [07:30 AM]), and the number of days that satisfy the condition (e.g., `5 days'). 
The system automatically updates the query result when people manipulate the time or data source on the screen, until they dismiss the query bar by swiping it to the left.
The design was inspired by the \textit{ambiguity widgets} from prior systems (e.g., \cite{Gao2015DataTone, Setlur2016Eviza, Srinivasan2018Orko, Srinivasan2020DataBreeze}).


\subsubsection{Speech and Touch Interaction}
\label{sec:mm-interaction}
Using touch and speech input modalities, Data@Hand provides three types of interaction---\textit{touch-only}, \textit{speech-only}, and \textit{touch+speech}.
With Data@Hand, people can directly interact with graphical widgets with touch, as they would normally do with any mobile apps. For example, they can tap on an item (e.g., label, chart) to select it, and swipe the range widget to shift the time frame back and forth.
Using speech, people can issue powerful commands that are applicable to a global scope: for example, changing a data source and date range together with \textit{``Show me the step counts from this summer,''} or executing a data-driven query with ``\textit{Highlight the days I walked more than 10,000 steps}.''
To handle speech, Data@Hand adopts a ``push-to-talk'' technique: the system records the speech input while people are pressing on the global microphone button~\symbolmic.
Given that pressing on the global microphone button serves only as a means to initiate speech interaction, we consider using speech with the global microphone button as speech-only interaction.

For the touch+speech interaction, people press on a target element while uttering speech commands that make use of the specific context the target element provides. 
For example, as shown in \circledigit{2} in \autoref{fig:teaser}, people can navigate to a different date range by uttering only a date (e.g., ``\textit{January 1}'') while pressing on the ``start date'' part of the range widget. 
Furthermore, with touch+speech people can perform a related command while keeping the same context. For example, as shown in \circledigit{3} in \autoref{fig:teaser}, after comparing the sleep ranges of winter and summer of this year (2020), they can simply utter ``\textit{Summer 2019}'' while pressing on the \chartplotname{} for the winter to compare summer of 2019 and 2020.
As demonstrated in previous research~\cite{vo1993multimodal,Martin1998TYCOON,Srinivasan2020InChorus}, this context helps Data@Hand facilitate faster interactions and reduce the complexity of speech commands, and ultimately improves user experiences.

\subsubsection{Speech Affordance and Feedback}

\begin{figure}[b]
    \centering
    
    \ifdim\textwidth=\columnwidth 
        \includegraphics[width= 0.5\textwidth]{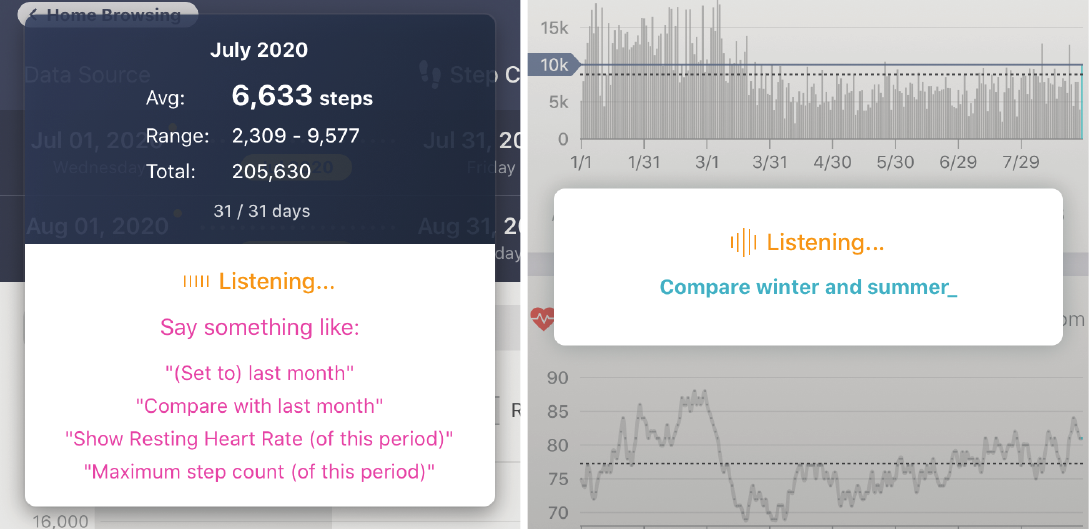}
    \else 
        \includegraphics[width=\columnwidth]{figures/speech_input_panel.pdf}
        \vspace{-4mm}
    \fi
    \caption{The speech input panel that displays preemptive guides (left) and a real time dictation result (right), before and during a speech interaction.} 
    \Description{Visual feedback of Data@Hand's speech input before and during a speech interaction.}
    \label{fig:feedback:input}
\end{figure}

When people press the global microphone button \symbolmic or long-press a target element for touch+speech interaction, Data@Hand displays the \textit{speech input panel} (\autoref{fig:feedback:input}) while dimming outside. If the target element is an \chartplotname{}, the speech input panel is embedded in a tooltip (\autoref{fig:feedback:input}, left). As a preemptive guide, Data@Hand displays a contextual message (e.g., ``Say a new date'' for start/end date) or example phrases on the panel (\autoref{fig:feedback:input}, left) until people start to utter. While people are talking to the system, the panel displays a real-time dictation result (\autoref{fig:feedback:input}, right) to make people aware of how their utterance is being recognized and to prevent the early release of the finger before the utterance is completely dictated. 

\begin{figure}[t]
    \centering
    
    \ifdim\textwidth=\columnwidth 
        \includegraphics[width= 0.5\textwidth]{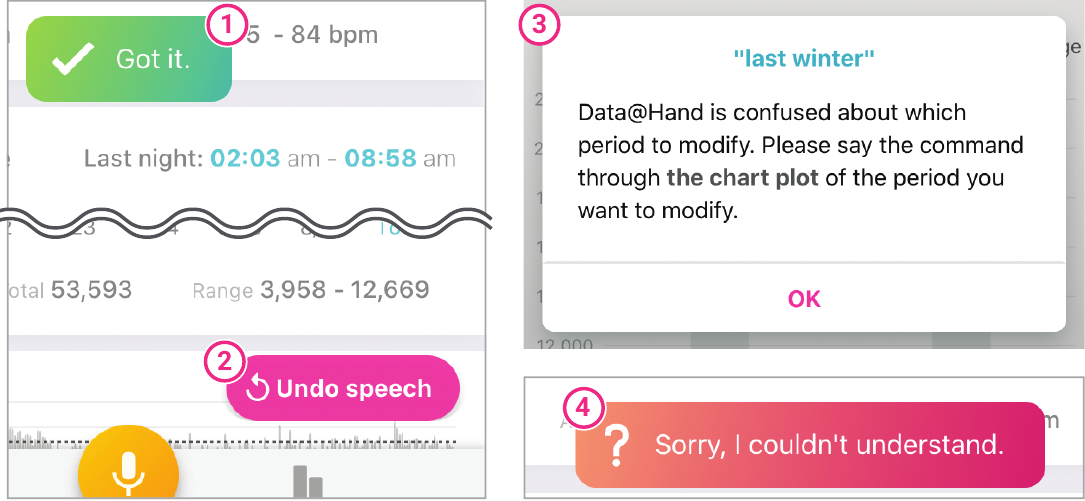}
    \else 
        \includegraphics[width=\columnwidth]{figures/speech_result_feedback.pdf}
        \vspace{-4mm}
    \fi
    
    \caption{When the system could translate the utterance and execute a valid operation, it momentarily displays a confirmation message~\circledigit{1} along with the undo button~\circledigit{2}.  When the translated operation is not valid, the system opens a contextual message dialog~\circledigit{3}. When the system fails to translate the utterance, it informs people accordingly~\circledigit{4}.} 
    \Description{Visual feedback of Data@Hand's speech input after a speech interaction.}
    \label{fig:feedback:result}

\end{figure}

After each execution of an operation, Data@Hand provides different types of feedback depending on its result. 
When Data@Hand could translate the utterance and execute a valid operation, it momentarily displays a confirmation message~(\circledigit{1} in \autoref{fig:feedback:result}) along with the undo button~(\circledigit{2} in \autoref{fig:feedback:result}) as a quick recovery option.
If the translated operation is invalid, the system opens a contextual message dialog~(\circledigit{3} in \autoref{fig:feedback:result}). For example, if one utters just one date range using speech-only interaction on the Two-range Comparison page, the system suggests try the same command using touch+speech interaction through the \chartplotname{} for disambiguation.
When Data@Hand fails to translate the utterance, it informs people accordingly~(\circledigit{4} in \autoref{fig:feedback:result}).  

\subsection{Interacting with Data@Hand}
Here we describe Data@Hand's interactions through a usage scenario: a self-tracker Zoe has been collecting data using a Fitbit band for almost five years. While this scenario emphasizes the speech-only and touch+speech interactions, most operations are also supported by touch-only interaction using graphical widgets. (Refer to our supplementary video to see more detailed interaction.)

\ipstart{Data Navigation \& Data-Driven Queries}
Being curious about her long-term activity patterns, Zoe opens Data@Hand. The system initially shows the past seven days of data on the Home page.
To extend the scope, she long-presses the start date and utters ``\textit{January~1}'' (\circledigit{2} in \autoref{fig:teaser}), and the system sets the date range to January 1, 2020 to today (August 27, 2020).
Skimming the bar chart for step counts, she notices a dramatic drop of step counts since mid-March, being reminded of the start of the COVID-19 lockdown. This plummet motivates her to explore her data since the lockdown has taken effect. 

Zoe starts to explore her step counts wanting to see how many days she achieved her daily step goal. She speaks \symbolmic``\textit{Days I met my step goal.}'' Referring to her step count goal (10,000 steps) from her Fitbit account, Data@Hand highlights the days with step counts higher than 10,000. She finds that the highlighted days are dense earlier in the year but sparse since March.
Zoe scrolls through the charts for other data sources. Once she reaches the chart for sleep range, she notices that her sleep has been pushed back since March.
Seeing this, Zoe decides to take a detailed look at her sleep ranges. To narrow down the scope to a more recent period, she speaks \symbolmic~``\textit{Sleep range of this month}.'' The system opens the Data Source Detail page for August's sleep range. By asking \symbolmic~``\textit{Days I woke up earlier than 7:30 AM},'' she learns that she woke up that early just for five days in August (\autoref{fig:scenario:detail}).

\ipstart{Temporal Comparisons}
Zoe is curious about how her sleep differs from last year's. She utters \symbolmic~``\textit{Compare with last August}'' (\circledigit{2}~in \autoref{fig:scenario}). Translating last August to \textit{August 2019}, Data@Hand opens the Two-range Comparison page comparing August 2019 against August 2020~(\autoref{fig:scenario:tworange:aug}). Zoe notices that this August's average sleep schedule is shifted by more than an hour late compared to August 2019. Also, the ranges of the bedtime and wake time in August 2020 are longer, implying her irregular sleep pattern. 
Zoe now wonders how the lockdown has affected her sleep. So, she changes the range from \textit{August 2019} to \textit{February 2020}, the previous month before the lockdown, by uttering ``\textit{February 2020}'' while long-pressing the \chartplotname{} for August 2019 (\circledigit{3}~in \autoref{fig:scenario}). She confirms that, compared to February, her sleep schedules for August are also shifted towards later hours and show more irregular bedtime and wake time (\autoref{fig:scenario:tworange:feb}).

To see how the lockdown affected her sleep schedules from the monthly trend, she speaks \symbolmic~``\textit{Show 2020 by month}'' (\circledigit{4} in \autoref{fig:scenario}). Data@Hand opens the Cyclical Comparison page, with her sleep ranges in 2020 grouped by month (\autoref{fig:scenario:cyclical}). She learns that her average sleep schedules have become more regular since May and that they have shifted to earlier hours.
Zoe continues on the exploration, switching to other data sources by swiping the \textit{data source widget} on the app header. 


\subsection{Implementation}
We implemented Data@Hand in TypeScript~\cite{TypeScript} upon React Native~\cite{ReactNative} to support both iOS and Android. 
When a participant first signs in with a Fitbit account through OAuth 2, the system uses Fitbit REST API~\cite{FitbitWebApi} to download the Fitbit data of the entire period since the account creation to the local SQLite database. The system always uses the locally-cached data to improve the performance by minimizing network overheads.

We used the Apple speech framework~\cite{AppleSpeech} and Microsoft Cognitive Services~\cite{MicrosoftCognitiveSpeech} as a speech-to-text recognizer on iOS and Android, respectively. 
We initially used a built-in speech recognizer for each OS. 
However, for Android, we decided to use Microsoft's Speech service because of the limitations of Android's built-in speech recognizer API.
We appended a set of application-specific keywords (e.g., name of the data sources) and time expressions (e.g., ``May'' is likely to refer a month rather than a verb) to the recognizers' vocabulary to improve accuracy for short phrases.

We implemented the system interpreter to work locally on the device. Receiving the recognized input text, the interpreter preprocesses it by performing part-of-speech tagging using the Compromise~\cite{Compromise} Javascript library and identifying parameters such as data sources, query conditions, and periods. To identify the time information mentioned in the input text, we used a customized version of Chrono~\cite{chrono}, a natural language time parsing library. After the preprocessing, the interpreter infers the operation based on the tagged verbs and parameters, the current screen information, and the pressed element for the touch+speech interaction.
\section{User Study}

\begin{table*}[t]
\makebox[\textwidth]{
\small\sffamily
			\def\arraystretch{1.3}
		    \setlength{\tabcolsep}{0.5em}
		    \centering
\begin{tabular}{|p{0.05\textwidth}ccp{0.28\textwidth}lp{0.14\textwidth}p{0.25\textwidth}|}
\hline
\rowcolor{tableheader} 
\textbf{Alias} & \textbf{Age} & \textbf{Gender} & \textbf{Occupation}            & \textbf{Fitbit usage} & \textbf{Fitbit wearable} & \textbf{Collected data}         \\
\hline
P1             & 27                                                       & M               & Client services trainer        & 2y 4m                 & Versa lite edition       & Step, HR, Sleep, Weight (Aria) \\
\arrayrulecolor{tablegrayline}\hline
P2             & 27                                                       & F               & Chemical engineer              & 4y 6m                 & Alta                     & Step, HR, Sleep                \\
\hline
P3             & 28                                                       & F               & Freelance social media manager & 4y 10m                & Versa                    & Step, HR, Sleep                \\
\hline
P4             & 32                                                       & M               & Implementation specialist      & 5y 1m                 & Ionic                    & Step, HR, Sleep, Weight (BT)   \\
\hline
P5             & 35                                                       & F               & Freelance photographer         & 5y 2m                 & Blaze                    & Step, HR, Sleep, Weight (BT)   \\
\hline
P6             & 23                                                       & F               & Graduate student               & 4y 8m                 & Alta HR                  & Step, HR, Sleep                \\
\hline
P7             & 23                                                       & F               & Graduate student               & 5y 6m                 & Charge 2                 & Step, HR, Sleep, Weight        \\
\hline
P8             & 46                                                       & F               & Product manager                & 1y 7m                 & Versa                    & Step, HR, Sleep, Weight        \\
\hline
P9             & 30                                                       & F               & Healthcare consultant          & 5y 9m                 & Versa                    & Step, HR, Sleep, Weight (Aria) \\
\hline
P10            & 24                                                       & M               & Unemployed (Varsity football player)                    & 1y 8m                 & Versa                    & Step, HR, Sleep, Weight        \\
\hline
P11            & 28                                                       & M               & Software engineer              & 4y 1m                 & Charge 3                 & Step, HR, Sleep, Weight        \\
\hline
P12            & 36                                                       & F               & Professional figure skater     & 2y 1m                 & Alta                     & Step, HR, Sleep, Weight        \\
\hline
P13            & 39                                                       & F               & Market survey manager          & 4y 4m                 & Alta                     & Step, HR, Sleep, Weight   \\
\arrayrulecolor{black}\hline
\end{tabular}
}

\footnotesize\begin{flushright} \textit{BT: Bluetooth scales which are from other vendors but can feed the data to Fitbit.}\end{flushright}

\caption{Summary of demographic and the Fitbit experience of our study participants.}
    \label{table:demographic}
\Description{Summary of demographic and the Fitbit experience of our 13 study participants.}
\vspace{-5mm}
\end{table*}

We conducted an exploratory study with Data@Hand, employing a think-aloud protocol, to examine how multimodal interactions help people explore their data. As part of this study, participants interacted with their own Fitbit data using their smartphones.
Due to the COVID-19 outbreak, all study sessions were held remotely using Zoom video call (in July 2020).
In Section~\ref{sec:study-setup}, we explain precautionary action we took to deliver a remote tutorial and to ensure close monitoring of the study session while mitigating potential privacy invasion.
We refined both the system and study protocol through six pilot sessions with Fitbit users recruited from Reddit. 
This study was approved by the Institutional Review Board of the University of Maryland at College Park.

\subsection{Participants}

We recruited 13 participants (P1--13; nine females and four males) from Reddit by advertising the study on the subreddits for job postings in the United States. Our inclusion criteria were adults who (1) are native English speakers; (2) have used Fitbit wearables for at least six months and tracked at least three of the following measures: step count, heart rate, sleep, and weight; (3) are interested in looking at their Fitbit data; (4) are currently using iPhone or Android; (5) have no visual, motor, or speech impairments; (6) have used voice recognition systems within six months with a generally positive experience; and (7) can understand simple charts.

The demographic and Fitbit usage information of our study participants is shown in \autoref{table:demographic}. Participants' ages ranged from 23 to 46 (\textit{avg} = 30.62). Ten participants were full-time employees, two were graduate students, and one was unemployed. At the time of the study, participants had used Fitbit for an average of 47 months based on their account creation date. Four participants had been tracking weight using Aria~\cite{Aria} or third-party Bluetooth scales (which share the data with Fitbit). The screen sizes of participants' smartphones ranged from 4.7 to 6.1 inches (eight participants used iPhones with a 4.7-inch screen). We offered a 30 USD Amazon gift card for their participation.

\subsection{Study Setup and Procedure}
\label{sec:study-setup}

\subsubsection{Pre-study Preparation}
A fully-remote study using participants' own data required us to take extra precautions, such as mitigating potential invasion of privacy from the use of participants' own smartphones, preparing the training material, and establishing robust audio- and video- recording methods. 
Furthermore, because Fitbit allowed only 150 API calls per hour per account, we had to prefetch the Fitbit data before the study session. (Immediately after the study session, we deleted the participants' Fitbit data from our server.)
To do so, we sent participants a link to a web page where they could sign an electronic consent form and fill out a pre-study questionnaire asking their Fitbit usage patterns and experiences of using voice assistant systems. 
After completing the questionnaire, the participants were asked to sign in with their Fitbit accounts so that our crawler could 
cumulatively download participants' entire Fitbit data. 
We also delivered the Data@Hand app to participants through TestFlight (iOS) or Google Play beta testing (Android) so that they could install the app on their phone. 

\subsubsection{Remote Study Session}
\begin{figure*}[t]
    \centering
    \includegraphics[width= 0.77\textwidth]{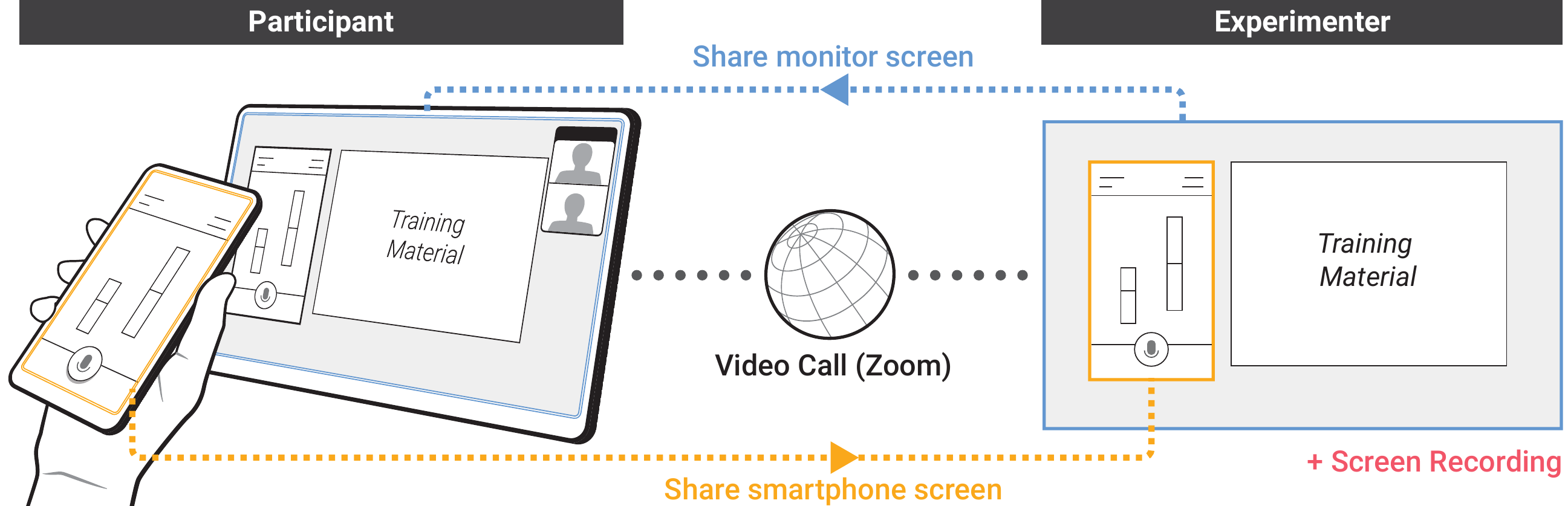}
    \Description{Illustration of two environments, one for a participant, the other for an experimenter. The two environments are connected via Zoom video call.}
    \caption{Settings of the remote study session using video call and screen sharing. The participant shares the smartphone screen with the experimenter, and the experimenter shares his monitor screen with the participant.}
    \label{fig:study_design}
\end{figure*}

Participants joined a 90-minute study session via Zoom video call~\cite{Zoom} from their computer. 
\autoref{fig:study_design} illustrates the settings of the remote study session. 
%
Via TeamViewer QuickSupport~\cite{TeamViewerQuickSupport}, participants shared their smartphone screen with the experimenter. Prior to the screen sharing, the experimenter instructed participants to remove any privacy-sensitive information from their home screen and to turn off all the notifications. 
The experimenter then shared his monitor screen with the participant using Zoom's screen sharing feature: the participant could see how their smartphone screen was being displayed to the experimenter. Using the recording feature in Zoom, the experimenter recorded the video call session including the shared smartphone screen. 

\ipstart{Tutorial}
After explaining the goal of the study, the experimenter gave a 40-minute tutorial, using an example dataset containing fabricated data generated based on the first author's four years of Fitbit data.
The tutorial covered Data@Hand's key design components and interactions---data sources \& charts, time manipulation, data navigation, temporal comparison, and data-driven queries. The experimenter introduced and demonstrated each feature using presentation slides and a video clip on a shared screen (refer to our supplementary material \revised{available at \weblinksupple}), and gave participants a chance to practice before moving to a new feature. In particular, participants were encouraged to practice the push-to-talk interaction exploring the example dataset. We gave them example speech queries (via shared Zoom screen) that they could use for practice, but also encouraged them to try out their own speech commands. We gave them enough time to practice until they feel comfortable using both the touch and speech input to interact with Data@Hand.

\ipstart{Free-form Exploration}
In this phase, the experimenter instructed participants to freely explore their own data with Data@Hand. 
We asked them to use \textit{any} of the supported interaction modalities (touch, speech, touch+speech) of their choice.
We also asked them to think aloud as they explored the data so that we can capture the insights they found and the challenges they faced, and understand their intentions and experiences. 
The experimenter observed how participants interacted with Data@Hand through the shared screen showing their smartphone screen and faces. 
We audio- and video-recorded each video call session and the system logged the interaction history and uploaded the logs to our server. This free-form exploration phase lasted approximately 20 minutes.

\ipstart{Debriefing}
We conducted a semi-structured interview around 10--15 minutes at the end of the session. We asked participants about their experiences with Data@Hand, difficulties and confusing features, and follow-up questions based on our observation in the exploration phase. We also asked them about the use cases of Data@Hand they could envision and their willingness to keep using the Data@Hand app after the study.

\subsection{Data Analysis}

We analyzed the video recordings and the interaction logs from the free-form exploration phase.
We performed both quantitative and qualitative analysis to examine how participants used the speech and touch modalities in finding personal insights.
As for the quantitative analysis, we first extracted participants' interaction attempts to perform actions for data navigation, temporal comparisons, and data-driven queries, reviewing the exploration videos and interactions logs. We defined an \textit{interaction attempt} as a series of low-level interactions (e.g., tapping, swiping) that were involved to obtain a desired outcome. To modify a start date, for example, people may tap the start date on the range widget to invoke a calendar popup and tap the target date. We treated this series of tap interactions as one attempt with touch-only interaction.  


As for the qualitative analysis, we analyzed the transcripts from the exploration phase to identify the types of personal insights, following Choe and colleagues' definition of \textit{personal insight} (``an individual observation about the data accompanied by visual evidence'')~\cite{Choe2017VisualizedSelf, Choe2015VisualizationInsights}. 
We extracted personal insights and categorized their types. 
For example, we extracted the following from P10's exploration session: (On the Cyclical Comparison page) ``\textit{Just pretty interesting that I get my most steps on Saturdays.}'' We coded this observation with three insight types: \textit{extreme} (``\textit{most steps}''), \textit{reference} (``\textit{Saturdays}''), and \textit{comparison by time segments} (essential to identify the day with the most step counts in this case). We describe when and how participants gained the insights in Section \ref{sec:res:insights}.

We transcribed the audio recordings of the debriefing interviews, which were grouped according to the following aspects: (1) participants' rationales of choosing the input modalities; (2) new analyses/tasks/questions Data@Hand enabled; (3) challenges participants encountered; and (4) how participants envisioned the use cases of Data@Hand in their own contexts.
We referenced this information when interpreting the video recordings and interaction logs, as well as to understand participants' general reactions to Data@Hand (reported in Section \ref{sec:res:reactions}).
\section{Results}
In this section, we report the results of our study in three parts: (1) interaction usages, (2) personal insights, and (3) general reactions to Data@Hand.

\subsection{Interaction Usages}
\label{sec:res:usage}

\begin{figure*}[b]
    \centering
    \includegraphics[width=\textwidth]{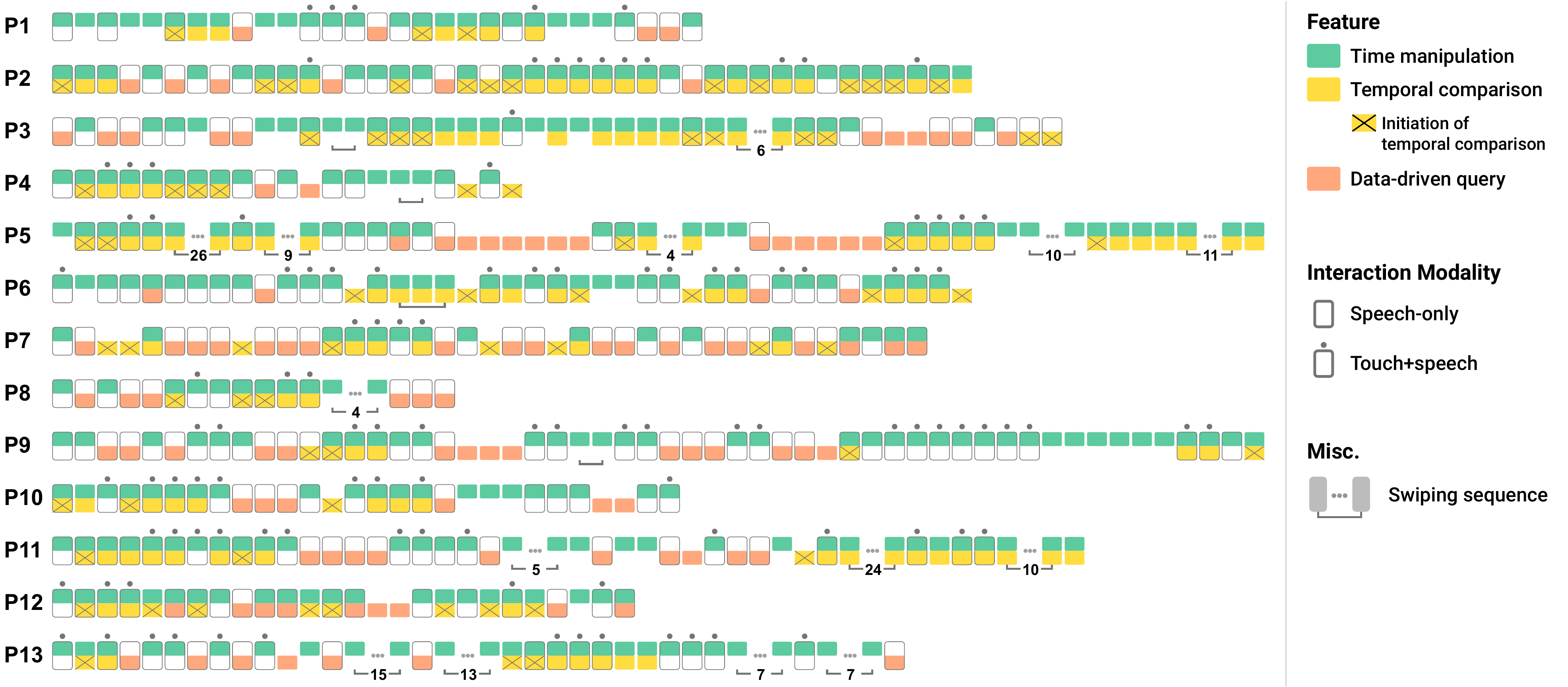}
    \vspace{-4mm}
    \caption{Sequences of operations---successfully executed interaction attempts---that are relevant to the three main features (time manipulation, temporal comparison, and data-driven query) by participant. Each unit on the horizontal-axis represents one operation, and the color of rectangles in a unit indicates the intended feature. The \textbf{X} mark indicates the initiation of temporal comparison (and thus only applicable for yellow rectangles; see \textbf{C1--4} in \autoref{tab:usage:comparison}). The border of a unit indicates the use of the speech modality (i.e., speech-only or touch+speech). 
    A series of swiping to manipulate time is bundled or collapsed with a black horizontal line. \revised{This operation overview shows that participants used all three interaction types to perform various actions. Also, the prevalence of a series of the green+yellow units without \textbf{X} suggests that participants often performed a series of comparisons with time refinement.}}
    \label{fig:timeline}
    \Description{A visualization of interaction history by participant.}
\end{figure*}

We identified 809 interaction attempts in total. Among these, 400 (49.4\%) were touch-only, 281 (34.7\%) were speech-only, and 128 (15.8\%) were touch+speech interaction attempts.
Among the 400 touch-only attempts, five were aborted by participants to perform the equivalent action using speech instead (e.g., P4 first opened a calendar picker to modify the start date, but he closed it and modified the start date using the touch+speech interaction).
Among the 281 speech-only attempts, 32 were failed due to the recognition/interpretation errors, 16 were invalid actions (e.g., \symbolmic~``\textit{Compare hours slept}'' without designating any comparative periods), and five were unsupported actions (e.g., attempting to execute a data-driven query to the aggregation plots).
Among the 128 touch+speech attempts, eight were recognition/interpretation errors, and seven were invalid actions (e.g., uttering a date where a period is required).
As a result, 736 (395 touch-only, 228 speech-only, and 113 touch+speech) interaction attempts were successfully executed, which we call \textit{operations} from now on. 
Of these, we included only 589 operations that are relevant to the three main features (time manipulation, temporal comparison, and data-driven queries) into further analysis (268 touch-only, 209 speech-only, and 112 touch+speech operations).
The rest 147 operations consisted of 127 touch-only and 19 speech-only operations for data source manipulation, and one touch+speech operation for data-driven query. 
\autoref{fig:timeline} visualizes these operations by participant. 
In the following subsections, we describe participants' detailed usage of these input modalities for the three main features.

\subsubsection{Time Manipulation}
\label{sec:res:usage:time}

\begin{table*}[t]
\footnotesize\sffamily
			\def\arraystretch{1.75}
		    \setlength{\tabcolsep}{0.2em}
		    \centering
\arrayrulecolor{black}

\begin{centering}

\arrayrulecolor[rgb]{0.8,0.8,0.8}
\begin{tabular}{!{\color{black}\vrule}p{0.13\textwidth}!{\color{black}\vrule}p{0.035\textwidth}p{0.31\textwidth}!{\color{lightgray}\vrule}P{0.032\textwidth}P{0.032\textwidth}!{\color{black}\vrule}c!{\color{black}\vrule}P{\participantcolumnwidth}!{\color{lightgray}\vrule}P{\participantcolumnwidth}!{\color{lightgray}\vrule}P{\participantcolumnwidth}!{\color{lightgray}\vrule}P{\participantcolumnwidth}!{\color{lightgray}\vrule}P{\participantcolumnwidth}!{\color{lightgray}\vrule}P{\participantcolumnwidth}!{\color{lightgray}\vrule}P{\participantcolumnwidth}!{\color{lightgray}\vrule}P{\participantcolumnwidth}!{\color{lightgray}\vrule}P{\participantcolumnwidth}!{\color{lightgray}\vrule}P{\participantcolumnwidth}!{\color{lightgray}\vrule}P{\participantcolumnwidth}!{\color{lightgray}\vrule}P{\participantcolumnwidth}!{\color{lightgray}\vrule}P{\participantcolumnwidth}!{\color{black}\vrule}} 

\arrayrulecolor{black}\hline
\rowcolor{tableheader} \textbf{Action} & \multicolumn{2}{l!{\color{lightgray}\vrule}}{\textbf{Operation Pattern}} & \multicolumn{2}{c!{\color{black}\vrule}}{\textbf{Modality}} & \textbf{Total} & \textbf{P1}                       & \textbf{P2}                        & \textbf{P3}                        & \textbf{P4}                        & \textbf{P5}                        & \textbf{P6}                        & \textbf{P7}                        & \textbf{P8}                        & \textbf{P9}                       & \textbf{P10}                       & \textbf{P11}                       & \textbf{P12}                       & \textbf{P13}                        \\
\arrayrulecolor{black}\thicktablehline
\multirow{5}{*}[2.8em]{\bfseries{\parbox[t]{3cm}{Modify\\time directly}}}                 & \patterncode{T1.} & Tap \lfbox[patternparam]{start/end date label} then pick \lfbox[patternparam, background-color=patterntime]{date}                    & \patterncode{T} & & 48                     & 11                                         & 1                                          & 13                                         & 1                                          & 6                                          & 3                                          &  &  & 3                                          & 3                                          & 4                                          & 1                                          & 2                                           \\ 
\arrayrulecolor[rgb]{0.8,0.8,0.8}\cline{2-19}
                                                 & \patterncode{T2.}  & Swipe \lfbox[patternparam]{range widget} until reaching the target                          &                                        \patterncode{T} &                                     & 170                    &  &  & 8                                          & 2                                          & 62                                         & 4                                          &  & 4                                          & 5                                          & 1                                          & 41                                         &  & 43                                          \\ 
\cline{2-19}
                                                 & \patterncode{T3.}  & \parbox[t]{7cm}{\symbolmictiny Speak <(\lfbox[patternparam, background-color=patterndatasource]{data source}) \lfbox[patternparam, background-color=patterntime]{period}>
                                                 \\
                                                 \patternexample{``\textit{\textcolor{patterntimetext}{Last 30 days}}'' /  ``\textit{\textcolor{patterndatasourcetext}{Step count} in \textcolor{patterntimetext}{2019}}''}
                                                 }                                & & \patterncode{S}                                           & 71                     & 6                                          & 11                                         & 5                                          & 6                                          & 5                                          & 7                                          & 8                                          & 3                                          & 7                                          & 6                                          & 4                                          & 3                                          &   \\ 
\cline{2-19}                    
 & \patterncode{T4.}  & \parbox[t]{7cm}{\symbolmulttiny \lfbox[patternparam]{start/end date label} and speak <\lfbox[patternparam, background-color=patterntime]{date}>
                                                 \\
                                                 \patternexample{\symbolmulttiny start date + ``\textit{\textcolor{patterntimetext}{January 1, 2019}}''}
                                                 }                           & \patterncode{T} & \patterncode{S}                                   & 75                     & 4                                          &  & 1                                          & 1                                          & 1                                          & 14                                         &  & 1                                          & 18                                         & 8                                          & 13                                         & 4                                          & 10                                          \\ 
\cline{2-19}
& \patterncode{T5.}  & \parbox[t]{7cm}{\symbolmulttiny \lfbox[patternparam]{\chartplotname} and speak <\lfbox[patternparam, background-color=patterntime]{period}>
                                                 \\
                                                 \patternexample{In the two-range comparison page, \\\symbolmulttiny the left period plot + ``\textit{\textcolor{patterntimetext}{January 2020}}''}
                                                 }                                 & \patterncode{T} & \patterncode{S}                                         & 37                     & 1 & 10                                         &  & 3                                          & 6                                          & 3                                          & 4                                          & 2                                          & 2                                          & 1                                          & 1                                          & 1                                          & 3                                           \\ 
\arrayrulecolor{black}\hline

\multirow{3}{*}[1em]{\bfseries\parbox[t]{3cm}{$^a$Execute\\comparison query}}
& \patterncode{C1+3.}  & Tap \lfbox[patternparam]{Compare button} then configure parameters   & \patterncode{T} & &                                                                            13                      & 1  &  & 1                                          &  & 1 & 2                                          &  &  & 1 &  &  & 5                                          & 2  \\ 
\arrayrulecolor[rgb]{0.8,0.8,0.8}\cline{2-19}
& \patterncode{C2.}  & \parbox[t]{7cm}{\symbolmictiny Speak <Compare (\lfbox[patternparam, background-color=patterndatasource]{date source}) \lfbox[patternparam, background-color=patterntime]{period} (\lfbox[patternparam, background-color=patterntime]{period})>
\\
                                                 \patternexample{``\textit{Compare \textcolor{patterntimetext}{January 2018} with \textcolor{patterntimetext}{January 2019}}''}
                                                 }                & & \patterncode{S}  & 30                     & 2                                          & 4                                          & 6                                          & 3                                          & 3                                          &  & 1                                          & 3                                          & 2                                          & 1  & 1                                          & 3                                          & 1                                           \\ 
\arrayrulecolor[rgb]{0.8,0.8,0.8}\cline{2-19}

& \patterncode{C4.}  & \parbox[t]{7cm}{\symbolmictiny Speak  <(\lfbox[patternparam, background-color=patterndatasource]{data Source}) \lfbox[patternparam, background-color=patterncycle]{cycle} \lfbox[patternparam, background-color=patterntime]{period}>
                                                 \\
                                                 \patternexample{``\textit{Show me \textcolor{patterndatasourcetext}{sleep} \textcolor{patterncycletext}{by month} for \textcolor{patterntimetext}{2020}}''}
                                                 }                        & & \patterncode{S}                                                                                    & 16                     &  & 8                                          & 3                                          & 1                                          & 1                                          &  &                                           &  & 1                                          & 1                                          & 1                                          &  &   \\

\arrayrulecolor{black}\hline

\multirow{1}{*}{\bfseries\parbox[t]{3cm}{\vspace{-1mm}$^b$Execute\\data-driven query}} & \patterncode{Q1.}  & \parbox[t]{7cm}{\symbolmictiny Speak <\lfbox[patternparam, background-color=patterncondition]{condition} \lfbox[patternparam, background-color=patterntime]{period}>
\\
\patternexample{``\textit{\textcolor{patternconditiontext}{Maximum step count} \textcolor{patterntimetext}{last month}}''}
}                                  & & \patterncode{S}                                & 10  & & & & & 1 & 1 & 3 & & & & & 5 & \\

\arrayrulecolor{black}\thicktablehline

\rowcolor[rgb]{0.937,0.937,0.937} \multicolumn{5}{|l|}{
    (): There exist operations which do not include this parameter.
}                                                                      & 470 & 25 & 34 & 37 & 17 & 86 & 34 & 16 & 13 & 39 & 21 & 65 & 22 & 61 \\
\hline
\end{tabular}
\end{centering}

\begin{flushleft}
            $^a$The occurrences are a subset of the operations of the equivalent patterns in \autoref{tab:usage:comparison}.\\
            $^b$The occurrences are a subset of the operations of the equivalent patterns in \autoref{tab:usage:query}.
\end{flushleft}

\vspace{1mm}
\caption{Summary of operations that contributed to the time manipulation, with the number of occurrences per participant and example utterances from participants. The modality column indicates the input modality of the operation (T: touch-only, S: speech-only, TS: touch+speech). \textbf{T1--5} indicate the operations to directly manipulate time, whereas \textbf{C1--4} and \textbf{Q1} indicate the operations to execute the comparison or data-driven queries with time parameters.
}
\label{tab:usage:manipulation}
\Description{This table summarizes operations that contributed to the time manipulation, with the number of occurrences per participant and example utterances from participants. The modality column indicates the input modality of the operation. T1 to T5 indicate the operations to directly manipulate time, whereas C1 to C4 and Q1 indicate the operations to execute the comparison or data-driven queries with time parameters.}
\vspace{-5mm}
\end{table*}

In total, participants manipulated time 470 times (see \autoref{tab:usage:manipulation} for the summary). 
Participants specified time (\textbf{T1--5}; e.g., change the start/end date) or manipulated time as part of executing comparison (\textbf{C1--4}) or data-driven queries (\textbf{Q1}). 
When manipulating time, participants actively used both speech and touch: to navigate to a new date range, participants used speech-only interaction 71 times (\textbf{T3}) and a calendar picker with touch 48 times (\textbf{T1}). To modify ranges in the Two-range Comparison page, participants tended to use touch+speech interaction on the \chartplotname{} instead of touch-only interaction. As shown in \autoref{tab:usage:comparison}-top, 12 participants used touch+speech 37~times (\textbf{T5}) while 3 participants used touch-only 13 times (\textbf{T1}).

When participants modified only start or end date of the date range, their behaviors differed depending on the distance between their target date and the currently selected one. If the target date was close to the original one (especially within the same month), they preferred a calendar picker~(\textbf{T1}) as it would require only a few taps. On the other hand, if the target date was far from the currently selected date (e.g., several months away), they preferred touch+speech interaction (\textbf{T4}) by long-pressing the date label and mentioning the target date.

Three participants (P5, P11, P13) heavily used swipe to modify the date range (\textbf{T2}, 146 out of 170 times). For example, starting from the Data Source Detail page for weight in year 2020, P13 swiped through the year 2016, skimming the trend of each year (see swiping sequences in \autoref{fig:timeline}).
The touch-only swipe is a quick way to navigate through using a preset date range. 

\begin{table*}[t]
\footnotesize\sffamily
			\def\arraystretch{1.75}
		    \setlength{\tabcolsep}{0.2em}
		    \centering
		    
            \begin{centering}
\arrayrulecolor{black}
\arrayrulecolor[rgb]{0.8,0.8,0.8}
\begin{tabular}{!{\color{black}\vrule}p{0.13\textwidth}!{\color{black}\vrule}p{0.025\textwidth}p{0.315\textwidth}!{\color{lightgray}\vrule}P{0.032\textwidth}P{0.032\textwidth}!{\color{black}\vrule}c!{\color{black}\vrule}P{\participantcolumnwidth}!{\color{lightgray}\vrule}P{\participantcolumnwidth}!{\color{lightgray}\vrule}P{\participantcolumnwidth}!{\color{lightgray}\vrule}P{\participantcolumnwidth}!{\color{lightgray}\vrule}P{\participantcolumnwidth}!{\color{lightgray}\vrule}P{\participantcolumnwidth}!{\color{lightgray}\vrule}P{\participantcolumnwidth}!{\color{lightgray}\vrule}P{\participantcolumnwidth}!{\color{lightgray}\vrule}P{\participantcolumnwidth}!{\color{lightgray}\vrule}P{\participantcolumnwidth}!{\color{lightgray}\vrule}P{\participantcolumnwidth}!{\color{lightgray}\vrule}P{\participantcolumnwidth}!{\color{lightgray}\vrule}P{\participantcolumnwidth}!{\color{black}\vrule}} 

\arrayrulecolor{black}\hline
\rowcolor{tableheader} \textbf{Action} & \multicolumn{2}{l!{\color{lightgray}\vrule}}{\textbf{Operation Pattern}} & \multicolumn{2}{c!{\color{black}\vrule}}{\textbf{Modality}} & \textbf{Total} & \textbf{P1}                       & \textbf{P2}                        & \textbf{P3}                        & \textbf{P4}                        & \textbf{P5}                        & \textbf{P6}                        & \textbf{P7}                        & \textbf{P8}                        & \textbf{P9}                       & \textbf{P10}                       & \textbf{P11}                       & \textbf{P12}                       & \textbf{P13}                        \\
\arrayrulecolor{black}\thicktablehline

\rowcolor{darkgray} \multicolumn{19}{l!{\color{black}\vrule}}{\textcolor{white}{\textbf{Two-range Comparison}}} \\

\multirow{2}{*}{\bfseries\parbox[t]{5cm}{Execute two-range\\comparison query}} & \patterncode{C1.}  & Tap \lfbox[patternparam]{Compare button} then configure \lfbox[patternparam, background-color=patterntime]{period} (\lfbox[patternparam, background-color=patterntime]{period})             & \patterncode{T} & &                                                                             4                      &  &  & 1                                          &  &  & 2                                          &  &  &  &  &  & 1                                          &            \\

\arrayrulecolor[rgb]{0.8,0.8,0.8}\cline{2-19}
                                                 & \patterncode{C2.}  & \symbolmictiny Speak <Compare (\lfbox[patternparam, background-color=patterndatasource]{date source}) \lfbox[patternparam, background-color=patterntime]{period} (\lfbox[patternparam, background-color=patterntime]{period})>
                                                                 & & \patterncode{S}  & 30                     & 2                                          & 4                                          & 6                                          & 3                                          & 3                                          &  & 1                                          & 3                                          & 2                                          & 1  & 1                                          & 3                                          & 1                                           \\ 

\arrayrulecolor{black}\hline
\multirow{2}{*}{\bfseries\parbox[t]{5cm}{$^a$Modify\\time directly}} & \patterncode{T1.} & Tap \lfbox[patternparam]{start/end date label} then pick \lfbox[patternparam, background-color=patterntime]{date}                    & \patterncode{T} & & 13                     & 3                                         &                                           & 8                                         &                                           &                                           &                                           &  &  &                                           &                                           &                                           &                                           & 2                                           \\ 
\arrayrulecolor[rgb]{0.8,0.8,0.8}\cline{2-19}
                                                 & \patterncode{T2.}  & Swipe \lfbox[patternparam]{range widget} until reaching the target                        &                                        \patterncode{T} &                                     & 45                    &  &  & 6                                          &                                           & 39                                         &                                           &  &                                           &                                           &                                           &                                          &  &                                           \\ 
\cline{2-19}
                                                 & \patterncode{T4.}  & \symbolmulttiny \lfbox[patternparam]{start/end date label} and speak <\lfbox[patternparam, background-color=patterntime]{date}>
                                                                            & \patterncode{T} & \patterncode{S}                                   & 5                     &                                           &  &                                           &                                           & 1                                          &                                          &  &                                           & 2                                         &                                           &                                         & 2                                          &                                           \\ 
\cline{2-19}
                                                 & \patterncode{T5.}  & \symbolmulttiny \lfbox[patternparam]{\chartplotname} and speak <\lfbox[patternparam, background-color=patterntime]{period}>
                                                                                  & \patterncode{T} & \patterncode{S}                                         & 37                     & 1 & 10                                         &  & 3                                          & 6                                          & 3                                          & 4                                          & 2                                          & 2                                          & 1                                          & 1                                          & 1                                          & 3                                           \\

\rowcolor{darkgray} \multicolumn{19}{l!{\color{black}\vrule}}{\textcolor{white}{\textbf{Cyclical Comparison}}} \\

\multirow{2}{*}{\bfseries\parbox[t]{5cm}{Execute cyclical\\comparison query}} & \patterncode{C3.}  & \parbox[t]{10cm}{Tap \lfbox[patternparam]{Compare button} then \\ configure (\lfbox[patternparam, background-color=patterndatasource]{data source}) \lfbox[patternparam, background-color=patterncycle]{cycle} (\lfbox[patternparam, background-color=patterntime]{period})} & \patterncode{T} & & 22                     & 1                                          &  &  & 2                                          & 1                                          & 4                                          & 5                                          &  & 1                                          & 1                                          & 1                                          & 4                                          & 2                                           \\ 
\arrayrulecolor[rgb]{0.8,0.8,0.8}\cline{2-19}             

  & \patterncode{C4.}  & \symbolmictiny Speak  <(\lfbox[patternparam, background-color=patterndatasource]{data Source}) \lfbox[patternparam, background-color=patterncycle]{cycle} (\lfbox[patternparam, background-color=patterntime]{period})>                     &  & \patterncode{S} & 19                     &  & 9                                          & 3                                          & 1                                          & 1                                          &  & 2                                          &  & 1                                          & 1                                          & 1                                          &  &   \\ 

\arrayrulecolor{black}\hline

\multirow{2}{*}{\bfseries\parbox[t]{5cm}{$^b$Modify\\time directly}} & \patterncode{T1.} & Tap \lfbox[patternparam]{start/end date label} then pick \lfbox[patternparam, background-color=patterntime]{date}                  & \patterncode{T} & & 3                     &                                          & 1                                          &                                          &                                           & 2                                          &                                           &  &  &                                           &                                           &                                           &                                           &              
\\ 
\arrayrulecolor[rgb]{0.8,0.8,0.8}\cline{2-19}
                                                 & \patterncode{T2.}  & Swipe \lfbox[patternparam]{range widget} until reaching the target                          &                                        \patterncode{T} &                                     & 54                    &  &  &                                           &                                           & 13                                         & 4                                          &  &                                           &                                           & 1                                           & 36                                         &  &                                           \\ 
\cline{2-19}
& \patterncode{T3.}  & \symbolmictiny Speak <\lfbox[patternparam, background-color=patterntime]{period}>                            & & \patterncode{S}                                           & 10                     & 1                                          & 3 &                                           &                                           &                                           &                                          & 3 &                        &                   & 1                                         & 2                                                                                   &                                           & \\
\cline{2-19}

                                                 & \patterncode{T4.}  & \symbolmulttiny \lfbox[patternparam]{start/end date label} and speak <\lfbox[patternparam, background-color=patterntime]{date}>                           & \patterncode{T} & \patterncode{S}                                   & 17                     &                                           &  &                                           &                                           &                                           &  5                                        &  &                                           &                                          & 4                                          & 7                                        &                                           & 1                                          \\ 

\arrayrulecolor{black}\thicktablehline
\rowcolor[rgb]{0.937,0.937,0.937} \multicolumn{5}{|l|}{(): There exist operations which do not include this parameter.}                                                                      & 259                    & 8                                         & 27                                         & 24                                         & 9                                         & 66                                        & 18                                         & 15                                         & 5                                         & 8                                         & 10                                         & 49                                         & 11                                         & 9                                          \\
\hline
\end{tabular}
\end{centering}
\begin{flushleft}
    $^a$$^,$$^b$ The occurrences are a subset of the operations of the equivalent patterns in \autoref{tab:usage:manipulation}.
\end{flushleft}
\vspace{1mm}
\caption{Summary of operations that contributed to establish a new comparison. \textbf{C1--4} indicate the operations to execute a comparison query and \textbf{T1--5} indicate the operations to directly manipulate time during the comparison, with their occurrences divided by the type of comparison.}
\vspace{-3mm}
\label{tab:usage:comparison}
\Description{This table summarizes operations that contributed to establish a new comparison information. C1 to C4 indicate the operations to execute a comparison query and T1 to T5 indicate the operations to directly manipulate time during the comparison, with their occurrences divided by the type of comparison.}
\end{table*}

\subsubsection{Temporal Comparisons}
\autoref{tab:usage:comparison} summarizes the operations to execute comparisons (\textbf{C1--4}), including the cases that modify time as part of the follow-up comparisons (\textbf{T1--5}).
As shown in \autoref{fig:timeline}, participants often performed a series of comparisons (184 operations, units in both green and yellow without an \textbf{X} mark) by refining the time range (or the data source).
When executing comparison queries (\textbf{C1--4}), participants tended to choose the input modalities depending on the type of comparison.
For \textbf{two-range comparison}, participants were inclined to use speech-only interaction (\textbf{C2}): only three participants used the Compare button to execute the two-range comparison query with touch-only interaction (\textbf{C1}, 4 instances).
On the other hand, participants showed mixed patterns on modality for \textbf{cyclical comparison}: among 12 participants who used the cyclical comparison, four participants (P1, P6, P12, and P13) used only touch~(\textbf{C3}), two (P2 and P3) used only speech~(\textbf{C4}), and the other six used both modalities.

\subsubsection{Data-Driven Queries}

\autoref{tab:usage:query} summarizes the operations dedicated to initializing and editing a data-driven query. Data@Hand supported only speech-only interaction to initialize a query~(\textbf{Q1}), with touch-only interaction only for follow-up editing of the recognized parameters through the widgets on the query bar~(\textbf{Q2}). In total, participants executed data-driven queries 89 times (\textit{avg} = 6.85).
Eight participants edited a recognized query parameter using a parameter widget 24 times. A majority of the data-driven queries were invoked to identify extreme values (e.g., lowest step count) or days when they achieved a goal (e.g., days with step counts higher than 10,000). However, participants also used data-driven queries to identify unusual days (e.g., days with step counts less than 3,000 steps for lazy days~[P5], days with bedtimes later than 5:00 AM for the days with sleep troubles~[P12]).

\renewcommand\participantcolumnwidth{0.027\textwidth}

\begin{table*}[t]
\footnotesize\sffamily
			\def\arraystretch{1.75}
		    \setlength{\tabcolsep}{0.2em}
		    \begin{center}
\arrayrulecolor{black}
\arrayrulecolor[rgb]{0.8,0.8,0.8}
\begin{tabular}{!{\color{black}\vrule}p{0.019\textwidth}p{0.42\textwidth}!{\color{lightgray}\vrule}P{0.032\textwidth}P{0.032\textwidth}!{\color{black}\vrule}c!{\color{black}\vrule}P{\participantcolumnwidth}!{\color{lightgray}\vrule}P{\participantcolumnwidth}!{\color{lightgray}\vrule}P{\participantcolumnwidth}!{\color{lightgray}\vrule}P{\participantcolumnwidth}!{\color{lightgray}\vrule}P{\participantcolumnwidth}!{\color{lightgray}\vrule}P{\participantcolumnwidth}!{\color{lightgray}\vrule}P{\participantcolumnwidth}!{\color{lightgray}\vrule}P{\participantcolumnwidth}!{\color{lightgray}\vrule}P{\participantcolumnwidth}!{\color{lightgray}\vrule}P{\participantcolumnwidth}!{\color{lightgray}\vrule}P{\participantcolumnwidth}!{\color{lightgray}\vrule}P{\participantcolumnwidth}!{\color{lightgray}\vrule}P{\participantcolumnwidth}!{\color{black}\vrule}} 

\arrayrulecolor{black}\hline
\rowcolor{tableheader}\multicolumn{2}{|l!{\color{lightgray}\vrule}}{\textbf{Operation Pattern}} & \multicolumn{2}{c!{\color{black}\vrule}}{\textbf{Modality}} & \textbf{Total} & \textbf{P1}                       & \textbf{P2}                        & \textbf{P3}                        & \textbf{P4}                        & \textbf{P5}                        & \textbf{P6}                        & \textbf{P7}                        & \textbf{P8}                        & \textbf{P9}                       & \textbf{P10}                       & \textbf{P11}                       & \textbf{P12}                       & \textbf{P13}                        \\
\arrayrulecolor{black}\thicktablehline

\patterncode{Q1.}  & \parbox[t]{7cm}{\symbolmictiny Speak <\lfbox[patternparam, background-color=patterncondition]{condition} (\lfbox[patternparam, background-color=patterntime]{period})>
\\
\patternexample{``\textit{Days I \textcolor{patternconditiontext}{slept more than six hours}}''}
}                                   & & S                                                                                    & 89                     & 4                                          & 6                                          & 9                                          & 1                                          & 3                                          & 4                                          & 19                                         & 6                                          & 11                                         & 4                                          & 9                                          & 7                                          & 6                                           \\ 
\arrayrulecolor[rgb]{0.8,0.8,0.8}\hline
                                                 \patterncode{Q2.}  & Tap \lfbox[patternparam]{parameter widget} then modify \lfbox[patternparam, background-color=patterncondition]{query parameter}  \vspace{0.7mm}                                             &                                    T &                                 & 24                     &  &  & 2                                          & 1                                          & 11                                         &  &  &  & 4                                          & 2                                          & 1                                          & 2                                          & 1                                           \\ \arrayrulecolor{black}\hline
\end{tabular}
\begin{flushleft}
                (): There exist operations which do not include this parameter.
\end{flushleft}
		    \end{center}
\vspace{1mm}
\caption{Summary of operations to execute data-driven queries. \textbf{Q1} indicates the operations to execute a query in natural language, and \textbf{Q2} indicates the operations to use the parameter widgets on the query bar to edit the recognized query.}
\vspace{-5mm}
\label{tab:usage:query}
\Description{This table summarizes the operations to execute data-driven queries. Q1 indicates the operations to execute a query in natural language, and Q2 indicates the operations to use the parameter widgets on the query bar to edit the recognized query. There are 89 operations for Q1, and 24 operations for Q2.}
\end{table*}

\begin{table*}[t]
\fontsize{7.5}{9.5}\sffamily
			\def\arraystretch{1.3}
		    \setlength{\tabcolsep}{0.5em}
		    \centering
\begin{tabular}{!{\color{black}\vrule}p{0.08\textwidth}p{0.135\textwidth}p{0.74\textwidth}!{\color{black}\vrule}}
\arrayrulecolor{black}\hline
\rowcolor{tableheader}
\multicolumn{2}{|l}{\textbf{Insight Type and Frequency}} & \textbf{Example Quotes} \\
\hline
\multirow{3}{*}{\parbox[t]{0.08\textwidth}{\textbf{Detail}\\(263)}}    & \parbox[t]{0.14\textwidth}{\textbf{Identify value}\\(109) - 12Ps} & ``\textit{What was that day? I got so much sleep. Wow, I got 13 and a half hours on that Monday.}'' -- P10 \\ \arrayrulecolor{lightgray}\cline{2-3}
 & \parbox[t]{4cm}{\textbf{Identify references}\\(104) - 13Ps}  & \parbox[t]{30cm}{``\textit{Um.. wonder why it's slow in April.}'' -- P2 
\\ 
``\textit{It's pretty consistent, but in December I had a very broad range the hours.}'' -- P6} \\\arrayrulecolor{lightgray}\cline{2-3}
 & \parbox[t]{4cm}{\textbf{Identify extreme}\\(50) - 12Ps}  & ``\textit{It looks like December was probably my highest activity month if I look back at the whole trend.}'' -- P8 \\ \arrayrulecolor{black}\hline

 \multirow{4}{*}{\parbox[t]{4cm}{\textbf{Comparison}\\(143)}} & \parbox[t]{4cm}{\textbf{Of two instances}\\ (66) - 11Ps}  & ``\textit{It's interesting to see my step count average is I was much more all over the place in October. But in February, I was much more consistent, which is better.}'' -- P1                                                                                                                                                                                                                    \\ 
\arrayrulecolor{lightgray}\cline{2-3}
 & \parbox[t]{4cm}{\textbf{By time segments}\\(43) - 10Ps}    & ``\textit{My average was cut about a half from January, February, down to March.}'' -- P4  \\ 
 
 \arrayrulecolor{lightgray}\cline{2-3}
  & \parbox[t]{4cm}{\textbf{By factor}\\(33) - 10Ps}  & ``\textit{My average (step count) was obviously higher six months ago than it is now because we're all locked in our houses.}'' -- P8 \\ 
 
 \arrayrulecolor{lightgray}\cline{2-3}
 
 & \parbox[t]{4cm}{\textbf{Against external data}\\(1) - 1P} & ``\textit{I've been tracking what I eat. So I definitely used to weigh more. Look at that. this is like 150 lb.}'' -- P11                                                                                                           \\ 
  \arrayrulecolor{black}\hline

\multirow{3}{*}{\parbox[t]{4cm}{\textbf{Recall}\\(115)}}     & \parbox[t]{4cm}{\textbf{External context}\\(79) - 13Ps} & 
``\textit{That{[}hourly step count chart{]} would remind me of what I did that day. I know, based on the time of day and the day of the week, that it was a hike that I went on and that's why I got the extra steps.}'' -- P8 \\ \arrayrulecolor{lightgray}\cline{2-3}
  & \parbox[t]{4cm}{\textbf{Confirmation}\\(26) - 8Ps} & ``\textit{I know I've been getting less sleep recently compared to before. That's what I wanted to see.}'' -- P11                                                                          \\ \arrayrulecolor{lightgray}\cline{2-3}
  & \parbox[t]{4cm}{\textbf{Contradiction}\\(10) - 5Ps}  & ``\textit{...because nothing feels the same (after the COVID-19 outbreak). But it's interesting to see that the data looks less terrible than I expected, and I'm kind of happy.}'' -- P12                                                          \\ \arrayrulecolor{black}\hline

\multicolumn{2}{|l}{\textbf{Value judgment} (51) - 10Ps}                               & ``\textit{My average bedtime is.. somewhere between 2 and 4 am..pretty terrible.}'' -- P9                                                                                                                                                                                                                                                                                                          \\ \hline
\multicolumn{2}{|l}{\textbf{Variability} (49) - 12Ps}                                  & ``\textit{Looks like I mean for everything in 2020, it seems to be my sleep is getting much more consistent, which is a good sign.}'' -- P1 \\ \hline
\multicolumn{2}{|l}{\textbf{Trend} (42) - 11Ps}                                         & ``\textit{It's an increasing trend but like it's super low in the beginning of April which I find odd.}'' -- P2 \\ \hline


\multicolumn{2}{|l}{\textbf{Correlation} (18) - 8Ps}                                   & \parbox[t]{30cm}{``\textit{The days that I have done the least amount of steps are the days of my heart rate is the lowest on average.}\\ \textit{That makes sense.}'' -- P3\vspace{0.4em}} \\ \hline
\multicolumn{2}{|l}{\textbf{Outlier} (7) - 3Ps}                                        & ``\textit{Wow, there is definitely an outlier there.}'' -- P1 \\ \hline
\multicolumn{2}{|l}{\textbf{Data summary} (6) - 3Ps}                                   & ``\textit{My average steps is 10760. Wow 4 millions (of total steps) *laugh* that's a lot. Range anywhere from eight to twenty four thousand, because sometimes I didn't wear it.}'' -- P6 \\           
 \hline
 
\end{tabular}
\vspace{2mm}
\caption{Visualization insight types identified from the transcripts from the exploration phase, frequency with the number of participants (Ps), and example quotes. The insight types are sorted by frequency.}
\label{tab:insights}
\Description{The table summarizes sixteen visualization insight types identified from the transcripts from the exploration phase, frequency with the number of participants, and example quotes. The insight types are sorted by frequency.}
\vspace{-3mm}
\end{table*}
\subsection{Personal Insights}
\label{sec:res:insights}

We extracted 367 data-driven observations (\textit{avg} = 28.23, \textit{min} = 10, \textit{max} = 52) from 13 participants and derived 694 personal insights. \autoref{tab:insights} summarizes the insight type categories, frequency, and example quotes for each category (refer to \cite{Choe2017VisualizedSelf} for definition of each category). Overall, participants gained various types of insights, covering most of those observed with a desktop personal data exploration tool~\cite{Choe2017VisualizedSelf} and from the data presentation videos of quantified-selfers (the enthusiastic self-trackers)~\cite{Choe2015VisualizationInsights}, except three categories: \textit{comparison by multiple services}, \textit{prediction}, and \textit{distribution by category}. 
In this section, we highlight notable categories and how Data@Hand supported gaining the insights.

Participants found 143 instances of \textbf{comparison} insights leveraging Data@Hand's two types of comparison: two-range comparison and cyclical comparison.
They actively drew existing knowledge, or \textbf{external contexts}, which were not captured in the data, in executing comparison queries; these contexts often served as a \textbf{factor} of comparison. For example, most participants were interested in comparing their activity level before and after the stay-at-home orders around mid-March 2020 caused by the COVID-19 outbreak; common patterns for this were to compare a recent month (after the lockdown) with the same month of previous years (e.g., July 2019 vs. July 2020), 
and to investigate the monthly \textbf{trend} of year 2020. 
Other factors participants considered include job changes, start of a new project, and school semesters. When participants were curious about a past period, they usually compared it with the recent period (e.g., ``\textit{This month},'' ``\textit{Last 90 days}'') as a reference.
  
Participants compared not only values (e.g., measurement values or their average) but also other aspects of their data, such as \textbf{trends} and \textbf{variability or consistency}.
In the comparison page, participants often inferred the variability from the aggregation plots, as each plot showed the range of the values (e.g., ``[Looking at the days-of-the-week comparison screen] \textit{...the time I went to bed was most inconsistent on Mondays, Wednesdays, and Saturdays.}'' -- P7). About a half of the variability instances (23 out of 49) were inferred from the \chartplotname{}s. When comparing \textit{trends} among different periods, however, participants used working memory, by sequentially navigating to each period because the comparison screens only provided with the aggregated information.

Eight participants sought \textbf{correlation} insights
by associating values from different data sources on the same day.
The highlights provided in response to data-driven queries served as visual links, as the same days were highlighted across all data sources. Guided by data-driven queries, participants often scrolled through the charts on the Home page or switched the data source on the Data Source Detail page, to identify similar patterns of peaks and drops against the superposed highlights.
For instance, P13 highlighted days with low resting heart rate (lower than 56) on the heart rate detail page, and then navigated to the step count page, finding the similarity between the highlighted days and the daily step count values on those days (``\textit{...the days my heart rate was low, was the days with my step count also, fairly low, for most days}'' -- P13).

\subsection{General Reactions to Data@Hand}
\label{sec:res:reactions}

In the debriefing interview, we gathered participants' feedback on Data@Hand.
In general, participants expressed excitement about the flexible time navigation and comparison features Data@Hand offered. Participants described time manipulation in natural language to be \textit{fast} and \textit{flexible}. For instance, P7 remarked, 
\begin{quote}
``\textit{I liked that I was able to change the date using the speech because I thought that was really easy rather than having to go through all of the different dates and months. And also it was cool to say like `around this date' or `this month' and it would get what you were talking about, whereas I think it could be hard to do in a traditional touch format.}'' 
\end{quote}
Nine participants contrasted their experience with Data@Hand to their previous experience with the Fitbit App, which was tedious to reach a specific date or period. P1 mentioned, ``\textit{If I was in the Fitbit App, I would have to go through a whole bunch of different screens to find one specific date and all of that data}.''
Participants also appreciated being able to view two periods side-by-side in the Two-range Comparison page. Seven participants reported that they wanted similar feature in the Fitbit App, with which they currently have to sequentially view the two periods, while relying on working memory. P3 remarked, ``\textit{It [Fitbit App] just isn't as efficient, but I definitely have gone and looked back. But it took so much time to do it, but by the time I got to the date from last year and then I got back to today, I forgot [the last year's value].}''


Lastly, to gauge the utility of Data@Hand beyond the study session, we asked participants if they would be willing to use Data@Hand in real life. (Data@Hand on their phones will continue to work by retrieving recent data directly from the Fitbit server.) All but one participant said that they would keep the Data@Hand app for continued use after the study. One participant decided not to keep the app because she had not been engaging in physical activities as usual due to the COVID-19 lockdown. 

\section{Discussion}
In this section, we discuss lessons learned from the design and implementation of Data@Hand as well as from the exploratory study. We also reflect on implications for better supporting personal data exploration through multimodal interaction in mobile contexts.

\subsection{Enhancing Personal Data Exploration through Flexible Time Manipulation}

Enabling flexible time manipulation was one of our key design rationales (\hyperref[sec:dr2]{DR2}), which we strongly believe facilitates the insights gaining process. We observed that Data@Hand's time manipulation feature was well received by participants, mainly due to the speech that enabled flexible expression of time that participants are already familiar with. 
To support time-based data exploration with personal data, a system needs a time range specified by start and end time points (this is commonly applied in typical graphical range widgets).
People, however, express time in a variety of ways, not necessarily specifying two time points. Examples include the range using the present date as an implicit anchor (e.g., ``\textit{Last six months}''), and the range presented with a semantic phrase (e.g., ``\textit{This March}'' indicating the range from March 1, 2020 to March 31, 2020). 
Furthermore, people may not remember the date for notable events (e.g., the summer solstice) and some events are not represented with a fixed date (e.g., Thanksgiving in the United States). Accommodating these diverse time expressions in Data@Hand enabled participants to easily and flexibly set time ranges. 

Major speech recognizers handle national holidays and seasons ingrained in our culture, but they do not know key personal events (e.g., job change, vacation, semester), which are important in exploring personal data. Thus, our study participants had to rely on their memory to perform time-based exploration using their personal contexts. Tagging personally meaningful events and being able to refer to them with speech could address this issue, offering great opportunities to enhance personal data exploration (e.g., \symbolmic ``\textit{Compare my sleep range between spring semester and summer break}'').

\revised{Although our work focused on a much-needed mobile scenario of exploring personal health data, we envision that multimodal interactions leveraging time expressions can be similarly useful in other contexts. For example, flexible time navigation and comparison features can facilitate exploring personal data beyond the health and fitness domains, such as productivity, finance, and location history (e.g., \symbolmic~``\textit{Compare last week's screen time with this week's},'' \symbolmic~``\textit{Most expensive expense during the last quarter},'' \symbolmic~``\textit{Show me the places that I visited in the last three months}'').}

\subsection{Complementary Roles of Speech and Touch Input Modalities}

Our quantitative results showed that participants used both speech and touch modalities, individually and in tandem, performing all three types of interactions---touch-only, speech-only, and touch+ speech.
Our observations and participants' feedback also suggest that participants made deliberate choices between the two input modalities. 
In the debriefing interview, participants distinguished the advantages of the two modalities. 

\textbf{Speech} interaction was generally considered to be fast and flexible, especially when making big shifts in terms of time ranges, or executing commands involving multiple keywords (e.g., \symbolmic ``\textit{Compare step counts of this month and last month,}'' from the Home page).
P13 remarked, ``\textit{I would like to be able to use voice for more things just because it's easier and I've found myself just sometimes thinking like, `Oh, would this be quicker if I use a voice command?' {`Would it be easier if I use a voice command?'}''} 
On the other hand, \textbf{touch} interaction was preferred in some cases, such as shifting time frames successively with swipe (see the swiping sequences from \autoref{fig:timeline}) or choosing a data source from a list with a tap. This is also reflected in the high number of touch-only operation for data source manipulation (127 out of 146). In addition, participants resorted to touch when they were having a difficulty remembering exact keywords for a speech command (e.g., ``month-of-the-year'' for the cyclical comparison).
Participants favored the \textbf{touch+speech} interaction when refining pre-executed commands (e.g., uttering a new date while pressing on the date button to modify only the start date). 
P4 noted, ``\textit{I felt much more confident doing that [touch+speech] because I knew that it was only got to manipulate that one aspect of a comparison chart or only the start date, instead of having to be more precise with my speech in what I was asking.}''

\subsection{Natural Language vs. Keyword-Based Commands}

We observed two different patterns of participants' use of speech commands---natural language (e.g., the two-range comparison and data-driven queries) and keyword-based utterances (e.g., uttering ``\textit{Hours slept}'' to set the data source). They impose different technical challenges.
The natural language commands were sensitive to the linguistic structure of participants' utterance. 
All 18 interpretation errors (the system recognized the utterance correctly but did not interpret the recognized text successfully) occurred during the natural phrasing of commands. To prevent such errors and improve the interpretation coverage of the system, we can collect utterance examples via crowdsourcing or from pilot studies to identify common linguistic structures people use to perform interactions.

On the other hand, keyword-based commands were vulnerable to recognition errors~\cite{Kim2019VoiceCuts} with the generalized speech-to-text recognizers we used.
All eight errors related to the keyword phrases occurred at the recognizer level. 
Such recognition errors may be prevented by training the recognizer with keywords as shown in recent projects \cite{Srinivasan2020InChorus, Srinivasan2020DataBreeze}. However, using a customized recognizer is currently not feasible for smartphone apps without involving a remote server, which may cause additional delays and thus hamper the user experience. 

\begin{revisedblock}
\subsection{Designing Multimodal Interaction for Smartphones}

Our main design goal with Data@Hand was to support the visual exploration of self-tracking data on smartphones. The smartphone form factor and personal data context led to the design choices that are different from general-purpose multimodal data exploration systems on tablet devices, such as InChorus~\cite{Srinivasan2020InChorus} and Valletto~\cite{Kassel2018Valletto}.

Multimodal interaction of InChorus and Valletto focuses on constructing visualization (e.g., performing data binding and visual encoding, specifying chart types) and their exploration is driven by attributes in a given table. For example, InChorus incorporates a wide range of multimodal interaction to support a flexible visualization construction on tablets: people can choose the input modality they prefer, such as drag-and-drop, point \& tap (using two fingers), point and write (with a pen) or speak, utter a command with speech, to perform a mapping between data attributes and chart elements, such as axes and legends. 
Valletto supports simple touch gestures such as swiping and rotating for manipulating visual encoding (e.g., flipping the axes) and provides a persistent chat panel where people can speak to generate a new chart, exclude/include attributes, or ask analytic questions such as correlation between two attributes.
However, it would be challenging to provide all of such interactions on a smartphone, which has a smaller screen, needs to be held with a non-dominant hand, and may not support a pen. 

Furthermore, in the personal data context especially to assist lay individuals, it is not necessary to support flexible visualization construction: it is more important to facilitate easy navigation and comparison across the time dimension in a given chart (e.g., bar chart and line chart). Therefore, Data@Hand's multimodal interaction focuses on the manipulation of time parameters, while reducing the complexity of interface. 

Our study results suggest that participants can learn and use such multimodal interaction to find various insights from their self-tracking data, and their reactions were generally positive. We believe that Data@Hand's multimodal interaction achieved a good balance between flexibility and learnability by carefully considering smartphone form factor and personal data context together. In addition, we note that, for the personal data exploration, Data@Hand's interaction can be transferred to tablet form factors.


\subsection{Privacy Concerns on Voice Interaction for Personal Data}
While describing their real-world use cases in the debriefing, seven participants noted that they would be inclined to use only touch in the public space for two main reasons: (1) they did not want to disturb others and (2) they were afraid that surrounding people might feel awkward seeing them verbalizing health-related queries. For example, P3 mentioned, ``\textit{If you're only able to use speech, there's no privacy. You can't be at the doctor's office and be like, `Tell me how much weight I've lost' or `What day was I the fattest last year.'}'' These remarks align with the findings from previous research that privacy concerns can discourage the use of voice interaction in close proximity to other people~\cite{EaswaraMoorthy2015PrivacyVoice}, thus potentially limiting the applicability of speech-incorporated multimodal interaction in the public setting.
\end{revisedblock}

\subsection{Improving Visual Representations}

According to our design rationale (\hyperref[sec:dr1]{DR1}), we used basic charts that many people are already familiar with, such as bar charts and line charts. 
We also designed a new representation (i.e., aggregation plots) to support temporal comparison, which requires data aggregations.
Furthermore, to efficiently communicate results for the data-driven queries, we enhanced basic charts with a highlighting capability while presenting data without aggregation (i.e., one bar/dot per day) in the Home page. 

We see opportunities to improve visual representations to further enhance data exploration experiences especially with long-term data. 
While participants could view year-long data without aggregation (which was shown to be readable in recent research~\cite{brehmer2018visualizing}), the current charts would not scale to view a longer term beyond a year.
One straightforward solution would be to use our aggregation plots with data grouping (e.g., by week, by month, by year), a common approach in existing mobile apps.
However, the lower level of details induced by the aggregation makes it difficult to highlight particular data points. It is an open research question to effectively show query results on these aggregation plots, for example, highlighting days with steps over 10,000 on the aggregation plots grouped by month.

\subsection{Conducting an Exploratory Study in a Remote Environment}
We had to address a number of challenges to convert the original plan of running the in-person study into a fully-remote one due to the COVID-19 outbreak. Here, we share some of the challenges and issues we encountered and how we alleviated them.
First, it was infeasible to effectively demonstrate multimodal interactions (e.g., the push-to-talk recording) during the remote tutorial. We therefore prepared a video clip with subtitles and played it during the tutorial to introduce interaction methods to our participants. 
Second, we did not have control over participants' environment. 
Before running the pilot sessions, we sent a checklist to our participants (e.g., turning off the smartphone notification, connecting the laptop to a power cable) and asked them to follow the instructions. However, some participants did not actually comply with the provisions, even if they confirmed that they did. Furthermore, participants were occasionally distracted with pets or family members. To alleviate such issues in the main study sessions, we thoroughly checked whether participants turned on the most strict do-not-disturb mode prior to the screen sharing. Also, when participants were interrupted, we paused the session and asked them to handle the situation (e.g., closing the room door). 

On the contrary, we were pleasantly surprised by unexpected advantages that our original in-person protocol would not have. First, because we were less constrained by time and location, we could reach a broad audience with diverse backgrounds and occupations. Second, the screen sharing app enabled us to record the smartphone screen in high-resolution, supporting better observations. Third, the research team members from remote locations could attend and observe the study session without much interference (by turning off the webcam and muting the microphone). 
We demonstrated that a remote study can be a viable option for deploying and testing multimodal interactions in a mobile app and hope that this study could inform other researchers wanting to design and run similar types of remote studies.

\section{Conclusion}


We presented Data@Hand, a novel mobile app that combines two complementary input modalities, speech and touch, to support exploring personal health data on smartphones. Data@Hand supports three types of interactions---touch-only, speech-only, and touch+speech---to enable flexible time manipulation, temporal comparisons, and data-driven queries.
To examine how multimodal interaction helps people explore their own data, we conducted an exploratory think-aloud study with 13 long-term Fitbit users. Participants successfully adopted multimodal interaction and used both speech and touch interactions while finding personal insights. We also learned when and why people choose one interaction modality over others. We highlighted several areas for future research, including incorporating personally meaningful events and contexts, improving the recognition and interpretation of speech commands, and refining visualizations for further enhancing data exploration. We also showed that a remote study can be a viable option for deploying and testing a mobile app with multimodal interaction.
In summary, our work contributes the first mobile app that leverages the synergy of speech and touch input modalities for personal data exploration, and the study conducted with participants' own long-term data on their devices.
We hope this work can inform and inspire researchers in the visualization and broad HCI communities to leverage multimodal interactions to foster fluid and flexible data exploration on smartphones.



\begin{acks}
\revised{
    We thank our study participants for their time and efforts. We also thank Niklas Elmqvist, Snehesh Shrestha, and the HCIL members at the University of Maryland at College Park, who provided feedback on the early version of this paper. Jarrett Lee, Lily Huang, Rachael Zehrung, Yuhan Luo, Pramod Chundury, and Ignacio Jauregui also helped us improve the tutorial protocol and material. This study was in part supported by the National Science Foundation award \#1753452. Young-Ho Kim was in part supported by Basic Science Research Program through the National Research Foundation in Korea, funded by the Ministry of Education (NRF-2019R1A6A3A12031352).
}

\end{acks}

\bibliographystyle{ACM-Reference-Format}
\bibliography{bibliography}

\end{document}